\documentclass[a4paper, 12pt]{article}

\usepackage{amsmath, amsfonts, amssymb, amsthm}
\usepackage[pdftex]{graphicx}
\usepackage{listings}
\usepackage{multirow}
\usepackage{placeins}
\usepackage{upgreek}
\usepackage[linesnumbered,ruled,vlined]{algorithm2e}
\usepackage{color}
\usepackage{lscape}
\usepackage{dsfont}
\usepackage[utf8]{inputenc}
\usepackage{natbib}
\usepackage{cancel}
\usepackage{comment}
\usepackage{enumerate}
\usepackage{appendix}
\usepackage[inline,shortlabels]{enumitem}
\usepackage{bm, bbm, dsfont}
\usepackage{authblk}
\usepackage[usenames,dvipsnames]{xcolor}
\usepackage[colorlinks=true,linkcolor=blue, citecolor=blue, urlcolor=blue]{hyperref}
\usepackage{float}
\usepackage{anyfontsize}
\usepackage{booktabs}
\usepackage{caption}
\usepackage[textfont=footnotesize,justification=centering]{subcaption}
\usepackage{soul}
\usepackage[multiple]{footmisc}
\usepackage{stackrel}
\usepackage[normalem]{ulem}
\usepackage{lineno} 
\usepackage{xurl}
\usepackage{csquotes}
\usepackage{longtable}
\usepackage{yfonts}
\usepackage{mathrsfs}

\bibpunct{(}{)}{;}{a}{,}{,}

\newcommand{\R}{\mathbb{R}}
\newcommand{\N}{\mathbb{N}}

\newcommand{\Eop}{\mathbb{E}}
\newcommand{\Pop}{\mathbb{P}}

\newcommand{\Pcond}{\mathbb{P}_{t,x, y, \ybaro}}

\newcommand{\FC}{\mathcal{F}}

\newcommand{\ybaro}{\overline{y}}

\newcommand{\xtil}{\tilde{x}}
\newcommand{\ytil}{\tilde{y}}
\newcommand{\rBrackets}[1]{\left( #1 \right)}
\newcommand{\sBrackets}[1]{\left[ #1 \right]}
\newcommand{\cBrackets}[1]{\left\{ #1 \right\}}
\newcommand{\tin}{t \in \mathcal{T}}

\newcommand{\TSet}{\mathcal{T}}

\newcommand{\Depi}{\mathcal{D}^{\widehat{\pi}}}

\newcommand{\Dbarepi}{\overline{\mathcal{D}}^{\widehat{\pi}}}

\newcommand{\XSet}{\mathcal{X}}
\newcommand{\YSet}{\mathcal{Y}}

\newcommand{\AMap}{\mathcal{A}}
\newcommand{\ASet}{\bm{\mathcal{A}}}
\newcommand{\epi}{\widehat{\pi}}
\newcommand{\gybar}[1]{g^{\ybaro}\left( #1 \right)}

\newcommand{\inttT}{\int \limits_{t}^{T}}

\newcommand{\intthT}{\int \limits_{t + \delta}^{T}}

\newcommand{\Bbar}{\overline{B}}
\newcommand\norm[1]{\left\lVert#1\right\rVert}

\definecolor{green}{rgb}{0.0, 0.5, 0.0}

\newtheorem{theorem}{Theorem}
\newtheorem{corollary}[theorem]{Corollary}
\newtheorem{lemma}[theorem]{Lemma}
\newtheorem{proposition}[theorem]{Proposition}
\theoremstyle{definition}

\newtheorem{remark}[theorem]{Remark}

\newtheorem{definition}[theorem]{Definition}
\numberwithin{equation}{section} \numberwithin{theorem}{section}

\def\0{\kern0pt\-\nobreak\hskip0pt\relax}

\makeatletter
\AtBeginDocument{%
\def\@serieslogo{%
\vbox to\headheight{%
\parindent\z@ \fontsize{6}{7\p@}\selectfont
\vss}}}

\def\makeoverbar#1#2#3#4#5#6#7{%
\setbox0=\hbox{$\m@th#2\mkern#5mu{{}#3{}}\mkern#6mu$}%
\setbox1=\null \dimen@=#4\fontdimen8#13 \dimen@=3.5\dimen@
\advance\dimen@ by \ht0 \dimen@=-#7\dimen@ \advance\dimen@ by \wd0
\ht1=\ht0 \dp1=\dp0 \wd1=\dimen@
\dimen@=\fontdimen8#13 \fontdimen8#13=#4\fontdimen8#13
\rlap{\hbox to \wd0{$\m@th\hss#2{\overline{\box1}}\mkern#5mu$}}
\fontdimen8#13=\dimen@}
\def\mylabel#1#2{{\def\@currentlabel{#2}\label{#1}}}

\makeatother

\DeclareFontFamily{LYG}{ygoth}{}
\DeclareFontShape{LYG}{ygoth}{m}{n}{<->ygoth}{}

\allowdisplaybreaks

\begin{document}

\title{Equilibrium investment under dynamic \\ preference uncertainty}

\date{}

\author{Luca De Gennaro Aquino\footnote{Department of Engineering, Reykjavik University, Iceland. \href{lucaa@ru.is}{lucaa@ru.is}}\quad  Sascha Desmettre\footnote{Institute of Financial Mathematics and Applied Number Theory, Johannes Kepler University Linz, Austria. \href{sascha.desmettre@jku.at}{sascha.desmettre@jku.at}} \\ Yevhen Havrylenko\footnote{Faculty of Business and Economics, University of Lausanne, Switzerland.  \href{yevhen.havrylenko@unil.cg}{yevhen.havrylenko@unil.ch}} \quad Mogens Steffensen\footnote{Department of Mathematical Sciences, University of Copenhagen, Denmark. {\href{mogens@math.ku.dk}{mogens@math.ku.dk}}} }

\maketitle

\vspace{-1cm}

\begin{abstract}
We study a continuous-time portfolio choice problem for an investor whose state-dependent preferences are determined by an exogenous factor that evolves as an It\^{o} diffusion process. Since risk attitudes at the end of the investment horizon are uncertain, terminal wealth is evaluated under a set of utility functions corresponding to all possible future preference states. These utilities are first converted into certainty equivalents at their respective levels of terminal risk aversion and then (nonlinearly) aggregated over the conditional distribution of future states, yielding an inherently time-inconsistent optimization criterion. We approach this problem by developing a general equilibrium framework for such state-dependent preferences and characterizing subgame-perfect equilibrium investment policies through an extended Hamilton\textendash Jacobi\textendash Bellman system. This system gives rise to a coupled nonlinear partial integro-differential equation for the value functions associated with each state. We then specialize the model to a tractable constant relative risk aversion specification in which the preference factor follows an arithmetic Brownian motion. In this setting, the equilibrium policy admits a semi-explicit representation that decomposes into a standard myopic demand and a novel preference-hedging component that captures incentives to hedge against anticipated changes in risk aversion. Numerical experiments illustrate how features of the preference dynamics --most notably the drift of the preference process and the correlation between preference shocks and asset returns--  jointly determine the sign and magnitude of the hedging demand and the evolution of the equilibrium risky investment over time.
\end{abstract}



\section{Introduction} \label{sec:intro}
\noindent

Optimal dynamic investment problems under uncertainty form a central theme in mathematical finance.
Classical formulations of these problems typically vary along three main dimensions: (i) the nature of the decisions to be made (investment, consumption, insurance, etc.); (ii) the structure of the underlying market (completeness, presence of jumps, stochastic market coefficients, and related features); and (iii) the objective functional (mean-variance trade-offs, expected utility maximization, constraints, or combinations thereof). Within this broad landscape, expected utility maximization remains the dominant paradigm, thanks to its well-established axiomatic foundation and compatibility with dynamic programming.

Most of the existing literature adopts a fixed parametric specification of preferences, typically power utility with a constant relative risk aversion (CRRA). While substantial effort has gone into modeling uncertainty in the financial market (e.g., by replacing deterministic returns and volatilities with stochastic processes), far less attention has been devoted to modeling uncertainty in preferences themselves. The prevailing assumption is that the decision maker knows her utility function and its parameters exactly. As a consequence, comparative statics with respect to the risk aversion coefficient are usually interpreted as comparisons across different agents, rather than as reflecting the uncertainty a single individual may face regarding her own risk attitudes. From a practical perspective, however, the choice of a utility function and its parameters is often the most contentious aspect of the modeling exercise, and skepticism about these inputs can undermine the normative force of the resulting optimization results.

This paper takes a different perspective by treating risk aversion itself as an uncertain and dynamically evolving quantity. Modeling such preference uncertainty introduces a number of conceptual choices. Even in the simple power utility setting, one must specify whether the risk aversion parameter is a random variable or a random process, whether it is observable or latent, what information about it can be learned over time, and whether it should be correlated with the financial market. Each of these modeling choices leads to a distinct dynamic optimization problem. In this paper, we focus on an investor who observes her current level of risk aversion, anticipates that it will evolve randomly over time, and takes this evolution into account when forming long-term investment plans. To capture this, we model risk aversion as a function of an observable diffusion, which may be correlated with the market, and thus can span cases where preferences and returns are independent as well as cases where they are systematically linked.

When formulating a decision problem under random risk aversion, a technical difficulty arises: outcomes evaluated under different utility functions are not directly comparable. Even for power utility, a payoff preferred at one risk aversion level may not be preferred at another, and utilities themselves live on incomparable scales. To resolve this, we map utilities associated with different future risk aversion levels onto a common scale by means of certainty equivalents, and then aggregate them through a flexible second-stage operator. The objective functional resulting from this normalization-aggregation procedure involves an integral of nonlinear transformations of conditional expectations, a structure known to generate time-inconsistency. As a consequence, the usual notion of optimal control is no longer appropriate. Instead, the proper solution concept is that of a time-consistent equilibrium strategy, meaning a strategy that is locally optimal at every point in time.

We develop a continuous-time framework for portfolio choice under evolving, state-dependent preferences and provide both the theoretical foundations and explicit characterizations of an equilibrium investment behavior in this setting.

Our first contribution is to formulate a coherent intertemporal criterion for a decision maker who anticipates that her future risk attitudes will change and that terminal wealth will ultimately be evaluated under the utility corresponding to the realized future state. As mentioned, to compare outcomes evaluated under different utilities, we normalize them through state-specific certainty equivalents and aggregate across future preference realizations using an outer evaluation function. This construction accommodates broad classes of utility functions and preference specifications, and it makes explicit how random, evolving risk aversion generates intrinsic time-inconsistency, even when preferences are fully observable and the market is otherwise standard.

Our second contribution is methodological. The equilibrium Hamilton\textendash Jacobi\textendash Bellman (HJB) formulations available in earlier studies cannot be applied directly to our framework: the objective involves conditioning on future preference states and an uncountable family of nonlinear transformations of conditional expected utilities. We develop an extension of the equilibrium HJB approach that addresses these two difficulties simultaneously. The key steps are: (i) deriving the dynamics of the state variables under conditioning on a future preference state (Section \ref{subsec:dynamics_under_different_measures}); and (ii) establishing the limiting form --as the number of approximation terms tends to infinity-- of the extended HJB system (eHJB) characterizing an equilibrium strategy for a finite-sum approximation of the original reward functional (Section \ref{subsec:heuristic_derivation_of_eHJB}). This yields a new system that captures the structure of preference uncertainty and is linked to several existing results on equilibrium investment under random risk aversion.

Our third contribution is a verification theorem showing that any solution to the eHJB indeed generates an equilibrium strategy (Section \ref{subsec:verification_results}). This result provides the conceptual bridge between the abstract equilibrium definition and the PDE characterization.

Finally, we apply our framework to a tractable CRRA specification in which the preference factor follows an arithmetic Brownian motion (Section \ref{sec:application}). In this setting, the equilibrium conditions reduce to a system that yields the semi-explicit representation of an equilibrium portfolio rule. The resulting policy decomposes into the familiar myopic demand and a new preference-hedging component that reflects the investor's incentive to adjust her exposure in anticipation of future changes in risk aversion. The structure, sign, and magnitude of this hedging term depend on the drift and volatility of the preference process, its correlation with asset returns, and the shape of the certainty equivalent aggregator. 

Even in this simplified CRRA environment, the equilibrium conditions give rise to a coupled nonlinear partial integro-differential equation (PIDE) for the auxiliary value functions associated with each potential preference state. This system cannot be solved analytically and presents significant numerical challenges due to its dimensionality and the continuum of conditioning arguments. To address this, we apply a neural network-based solution method that formulates the PIDE system as a physics-informed learning problem.

Numerical experiments based on this approach illustrate how preference dynamics shape both the hedging demand and the evolution of the equilibrium risky investment over time.

\medskip

\paragraph{Related literature.} Our work relates to and expands several research areas. Optimal investment under state-dependent utility has been studied in \cite{BVY18}, who adopt the martingale method to obtain explicit solutions and avoid issues of time-inconsistency by not using certainty equivalents for normalization. In their framework, random preferences are fully correlated with the financial market. Optimal consumption, investment, and insurance under state-dependent risk aversion in the health dimension are analyzed in \cite{SoS23}. As in \cite{BVY18}, they do not use certainty equivalents, but, in contrast, their random preferences are not correlated with asset returns.

Beyond these contributions, several papers consider time-varying risk attitudes more generally, including \cite{Netzer2009:AER}, \cite{Schildberg2018:JEP}, and \cite{Bekaert2022:MS}, who document how attitudes toward risk may adjust with economic conditions, learning, or endogenous feedback mechanisms. In addition, a substantial body of empirical evidence suggests that preferences themselves are uncertain and subject to latent heterogeneity. Experimental studies, such as \cite{WeberMilliman1997:MS}, \cite{Fischer2000:MS}, \cite{AndersenEtAl2008:IER}, and \cite{Brunnermeier2008:AER}, show that individuals display significant variation in measured risk aversion across tasks, contexts, and time, supporting the view that preferences may evolve with economic or personal circumstances.

A paper that is particularly close in spirit to ours is \cite{DesmettreSteffensen2023:MF}, who average over the distribution of certainty equivalents associated with different realizations of an individual’s risk aversion. The primary distinction is that they treat risk aversion as a static random variable, about which the decision maker receives no new information. In contrast, we model risk aversion as an observable stochastic process driven by a stochastic differential equation (SDE), thus offering a greater level of flexibility and dynamic structure.

 \cite{BS21} also employ certainty equivalents to normalize across a heterogeneous set of agents with varying preferences. More specifically, their objective is formulated as a two-stage utility functional, where an outer utility function is applied to the distribution of the agents’ certainty equivalents. Another related work is \cite{ChenGuanLiang2025}, in which risk aversion is determined by a finite-state Markov chain that identifies market regimes. Each regime determines both the drift and volatility of returns and the level of risk aversion, and the optimization problem aggregates expected certainty equivalents across regimes, leading to time-inconsistency.

On the methodological side, our approach is rooted in the equilibrium concept for time-inconsistent control problems. The interpretation that dynamically inconsistent preferences can be treated as a non-cooperative game between successive selves goes back to \cite{Strotz1956:RES}. A precise mathematical formalization in continuous time was provided by \cite{EkelandLazrak2010:MFE}, \cite{EkelandPirvu2008:MFE}, and \cite{EkelandMbodjiPirvu2012:SIFIN}, primarily in the context of non-exponential discounting. The mean-variance optimization problem, initially incorporated in this framework by \cite{BasakChabakauri2010:RFS}, was subsequently formalized in the general equilibrium approach by \cite{BjorkMurgoci2014:FS}, \cite{BjorkMurgociZhou2014:MF}, and \cite{BjorkKhapkoMurgoci2017:FS}. \cite{KrygerNordfangSteffensen2020:MMRO} provide a survey-style overview of objectives where time-inconsistency originates from nonlinearities such as the square function. A comprehensive review of time-inconsistent control theory with applications in finance is given in \cite{BjoerkKhapkoMurgoci2021:TICT}. Problems in which time-inconsistency arises from certainty equivalents have also been approached through equilibrium theory; see, for example, \cite{JensenSteffensen2015:IME} and \cite{FahrenwaldtJensenSteffensen2020:JME}, who aggregate certainty equivalents to disentangle time and risk preferences. 

The structure of our objective functional also bears a formal resemblance to the smooth ambiguity model of \cite{KlibanoffMarinacciMukerji2005:Econometrica, KlibanoffMarinacciMukerji2009:JET}, which separates attitudes toward risk from attitudes toward model uncertainty. However, the rationale is fundamentally different, as in our model, the aggregation reflects uncertainty about future preferences rather than ambiguity about probability models; we expand on this parallel in Remark \ref{remark:smooth_ambiguity}. Similar themes of model uncertainty in dynamic equilibrium problems can be found in \cite{GuanLiangXia2025:MOR}, who combine smooth ambiguity preferences with equilibrium strategies and learning about uncertain asset drift.

Finally, our setting is related to the literature on forward performance processes, as established by \cite{MusielaZariphopoulou2007investment,MusielaZariphopoulou2008optimal} and \cite{Z09}. They introduce a new class of dynamic utilities generated forward in time and allow these utilities to be stochastic processes, similar to the exposition in this paper. The key difference is that they do not rely on certainty equivalents, and time-inconsistency does not arise. In that direction, \cite{Maggis2025} more recently elaborated on the consistency of optimal portfolio choice for state-dependent exponential utilities, and found that a unique forward prediction of random risk aversion exists, ensuring the consistency of optimal strategies across any time horizon.

\paragraph{Structure of the paper} The rest of the paper unfolds as follows. Section \ref{sec:problem_formulation} introduces the economic environment, including the financial market, the preference state process, and the reward functional used to evaluate portfolio strategies. Section \ref{sec:derivation_equilibrium} develops the extended equilibrium HJB system and contains the verification theorem. Section \ref{sec:application} specializes the framework to a tractable CRRA setting. Section \ref{sec:Conclusions} concludes. All technical proofs and auxiliary results are collected in the Appendices. 

\section{Problem formulation}
\label{sec:problem_formulation} \noindent 

Let $(\Omega, \FC, (\FC_t)_{ 0\le t\le T}, \Pop)$ be a filtered probability space satisfying the usual conditions, where $T>0$ is a fixed time horizon, and let $B^1$ and $B^2$ be two independent standard Brownian motions on this space. We denote by $\TSet := [0, T]$ the investment period. 

The investor operates under a classical Black\textendash Scholes financial market with a risk-free asset and one risky asset:
\begin{align*} 
d S^0_t &= S^0_t r dt\,,\\
d S^1_t &= S^1_t \left( \mu_S dt + \sigma_S \rBrackets{ \rho d B^1_t + \sqrt{1 - \rho^2} d B^2_t} \right)\,,
\end{align*}
with $r, \mu_S \in \R, \sigma_S > 0$, $\rho \in [-1,1]$ constants. Here, we introduced two Brownian motions to keep track of two distinct sources of uncertainty. One of them, $B^1$, will also drive the evolution of the preference factor (described below) so that shocks to the risky asset can be correlated with shocks to risk aversion. The second, $B^2$, provides an independent source of randomness. 

In what follows, we denote by $\pi(t, x, y) \in \R$ the fraction of wealth invested in the stock $S^1$ at time~$t$, given current wealth $x$ and preference state $y$. The process $\pi = \rBrackets{\pi(t, X^\pi_t, Y_t)}_{\tin}$ is referred to as the portfolio strategy, the investment strategy, or simply the control. For brevity, we will often write $\pi(t)$ in place of $\pi(t, X^\pi_t, Y_t)$. (We adopt the notation $\pi(t)$, instead of $\pi_t$, to emphasize that the control depends explicitly on the current time --and state variables--, rather than to suggest a dynamic process indexed by $t$.)

The controlled wealth process $X^\pi = \rBrackets{X^\pi_t}_{\tin}$, under an admissible portfolio strategy $\pi$, is given by the solution of the SDE
\begin{equation}
\label{eq:SDE_X}
\begin{split}
 d X^\pi_t & = X^\pi_t (r+ \pi(t) (\mu_S-r))dt + X^\pi_t\pi(t) \sigma_S \rBrackets{ \rho d B^1_t + \sqrt{1 - \rho^2} d B^2_t}\,,\\
X^{\pi}_0 & =x_0 > 0.
\end{split}
\end{equation}
The drift  $\mu_X$ and diffusion $\sigma_X$ of $X^{\pi}$, in accordance with \eqref{eq:SDE_X}, may be written compactly as
\begin{equation*}
d X^\pi_t = \mu_X (t, X^{\pi}_t, \pi(t))dt + \sigma_X(t,X^{\pi}_t, \pi(t))dB_t,
\end{equation*}
where $B := (B^1, B^2)$. The wealth process takes values in $\mathcal{X}:=(0, \infty)$.

To model time variation in risk attitudes, we introduce an exogenous factor process $Y$, taking values in a state space $\YSet$, and governed by the diffusion process
\begin{equation}\label{eq:SDE_Y}
d Y_t = \mu_Y(t,Y_t) dt  + \sigma_Y(t,Y_t) dB^1_t, \qquad Y_0= y_0 \in \R,
\end{equation}
where  $\mu_Y, \sigma_Y: \TSet \times \YSet \mapsto \R$ are continuous functions ensuring that \eqref{eq:SDE_Y} has a unique strong solution and $\sigma_Y(t, y) \neq 0$ for all $(t,y) \in \TSet \times \YSet$. Note that while \cite{ChenGuanLiang2025} assume the independence between the preference factor and the risky asset, we allow for arbitrary correlation through the common Brownian motion $B^1$.

The process $Y$ serves as the sole driver of intertemporal fluctuations in preferences; that is, the investor's risk attitude at time $t$ is captured through the mapping $\gamma: \YSet \mapsto \mathbb{R}$, which parametrizes the curvature of the instantaneous utility function. Therefore, all changes in risk aversion arise from the stochastic evolution of $Y$.

Intuitively, the factor $Y$ may represent broad economic indicators, such as stock prices or volatility, or more individual circumstances, such as health conditions or habits. For a fixed realization of $Y_{T} = \bar{y}$, the value $\gamma(\bar{y})$ is then inserted into a utility specification, where, depending on the chosen utility family, it parametrizes either relative risk aversion or absolute risk aversion. (In the present formulation, only the terminal value $Y_T$ enters the utility evaluation, so we only require a parameterization of $\gamma$ at time $T$. Nonetheless, if one were to introduce intermediate consumption, the same mechanism would naturally extend to a time-varying risk aversion index $\gamma(Y_t)$, applied at each consumption time.)

The usual notions of constant relative or absolute risk aversion refer primarily to the parameter’s independence on wealth, and this feature is preserved here. What we allow, however, is that risk aversion may shift in response to the evolution of the underlying state variable $Y$. Thus, along the wealth dimension, the investor behaves like a standard CRRA or CARA agent, while economic or personal conditions summarized by $Y$ can make the investor effectively more or less risk-averse over time.

For each possible future preference state $\bar{y} \in \YSet$, we denote by $u^{\gamma(\bar{y})}: \XSet \to \mathbb{R}$ the von-Neumann-Morgenstern utility function associated with the risk aversion level $\gamma(\bar{y})$. We assume that each $u^{\gamma(\bar{y})}$ is increasing, strictly concave, and twice continuously differentiable, with strictly nonvanishing marginal utility, i.e., $(u^{\gamma(\bar{y})})'(x) \neq 0,$ for all $x \in \XSet$.

\begin{remark}
    The pair $(Y,\gamma)$ is intentionally not uniquely determined. Only the composite quantity $\gamma(\bar{y})$ matters for preferences, and different choices of $\bar{y}$ and $\gamma$ can produce the same effective specification. For example, in the numerical section, we formalize the state variable as an arithmetic Brownian motion and take $\gamma(\bar{y})=\exp(\bar{y})$, which leads to the risk aversion becoming a geometric Brownian motion. 
    The very same structure could instead be obtained by choosing $Y$ directly as a geometric Brownian motion and letting $\gamma$ be the identity function. More generally, any monotone reparameterization of $Y$, combined with the corresponding inverse adjustment of $\gamma$, leaves the induced preferences unchanged.

This apparent ambiguity is a feature rather than a limitation: it enables us to place different \emph{types} of risk aversion, such as relative and absolute risk aversion, within a common framework, in the sense that the same underlying factor $Y$, be that a stock price or a health index, may drive both the CRRA and the CARA. The model, therefore, accommodates situations in which several facets of risk attitudes respond to the same economic or personal conditions. 
\end{remark}

\medskip

Before defining the objective functional, we describe the utility structure induced by preference uncertainty. As mentioned, at the investment horizon, terminal wealth is evaluated under the utility function $u^{\gamma(\bar{y})}$, for each $\bar{y} \in \YSet$. However, utilities arising from different preference states are not directly comparable, hence after computing the conditional expected utility under the scenario $Y_T = \bar{y},$ 
we normalize it via the inverse utility $\left( u^{\gamma(\bar{y})}\right)^{-1}$, obtaining the certainty equivalent at the preference state $\bar{y}$:
$$\left(u^{\gamma(Y_T)}\right)^{-1}\Big(\Eop_{t,x,y, \ybaro}\left[u^{\gamma(Y_T)}(X^{\pi}_T)\right]\Big),$$
with the conditional expectation meant as $    \Eop_{t, x, y, \ybaro} \big[\cdot\big]
:=\Eop\big[\cdot \, \vert \, X^{\pi}_t = x, Y_t = y, Y_T = \ybaro \big]. $
These state-dependent certainty equivalents are then aggregated across all possible future preference states through a function $v : \XSet \to \mathbb{R}$, which is assumed to be increasing and twice continuously differentiable.

This produces the reward functional
\begin{equation}
\label{eq:reward_functional}
J^\pi(t,x,y) := \Eop_{t,x, y}\left[ v \circ \left(u^{\gamma(Y_T)}\right)^{-1}\Big(\Eop_{t,x,y, \ybaro}\left[u^{\gamma(Y_T)}(X^{\pi}_T)\right]\Big)  \right]\,, \\[0.2cm]
\end{equation}
where $v \circ \left(u^{\gamma(Y_T)}\right)^{-1}$ denotes the composition of $v$ and $\left(u^{\gamma(Y_T)}\right)^{-1}$.

This two-stage normalization-aggregation structure provides a coherent way to compare utilities generated under different future preference states, but it also introduces nonlinear conditioning on both the present and future preference states, thereby generating time-inconsistency.

\begin{remark} \label{remark:smooth_ambiguity}
The outer function $v$ in \eqref{eq:reward_functional} plays a role reminiscent of the second-order utility (also referred to as the ambiguity index) in the smooth ambiguity model of \cite{KlibanoffMarinacciMukerji2005:Econometrica}. In their model, Klibanoff, Marinacci, and Mukerji (KMM) aggregate expected first-order utility via a fixed function $\phi$, whose concavity (convexity) directly encodes ambiguity aversion (ambiguity loving). In our setting, aggregation occurs through the family of functions $v \circ \left(u^{\gamma(\ybaro)}\right)^{-1}$, which depend explicitly on the future preference state $\ybaro$ and are therefore random rather than fixed. 

Alternatively, one may regard $v$ itself as a deterministic aggregator applied to certainty equivalents (instead of expected utilities, as in the KMM model), conditional on $Y_T = \ybaro$. The curvature of $v$ therefore governs the attitude toward ambiguity in certainty equivalents, which is not directly analogous to the curvature of $\phi$. Therefore, the economic meaning of concavity differs across the two frameworks: while the concavity of $\phi$ captures aversion to model uncertainty, the concavity of $v$ captures aversion to dispersion in normalized payoffs arising from uncertain future preferences.

\end{remark}

We next specify the class of admissible investment strategies. Admissibility here requires that the wealth and preference processes remain well-defined under the chosen control, and that the reward functional $J^{\pi}(t,x,y)$ is finite for all initial states.
\begin{definition}[Admissible control law]\label{def:admissibility_new}
An admissible control law is a map $\pi : \TSet \times \XSet \times \YSet \to \R$ satisfying the following conditions:
\begin{enumerate}
\item For each initial point $(t,x, y) \in \TSet \times \XSet \times \YSet$, the SDEs \eqref{eq:SDE_X}-\eqref{eq:SDE_Y}
have a unique strong solution denoted by $X^\pi$, $Y$.
\item For each point $(t,x,y) \in \TSet \times \XSet \times \YSet$, we have
\begin{equation*}
\Eop_{t,x,y}\left[ v \circ \left(u^{\gamma(Y_T)}\right)^{-1}\left(\Eop_{t,x,y, \ybaro}\left[u^{\gamma(Y_T)}(X^{\pi}_T)\right]\right)  \right] < \infty \,.
\end{equation*}
\item $\pi$ is continuous in $t, x, y$.  
\end{enumerate}
The set of admissible strategies is denoted by $\ASet$. 
\end{definition}

Because the objective functional \eqref{eq:reward_functional} is time-inconsistent, one cannot rely on the dynamic programming principle to find optimal controls. Therefore, we seek to determine equilibrium control laws in the sense of the following definition. 
\begin{definition}[Equilibrium control law; cf. Def. 15.3 in \cite{BjoerkKhapkoMurgoci2021:TICT}]\label{def:equi_control}
Consider an admissible control law $\epi$ (informally viewed as a candidate equilibrium law). Choose an arbitrary $\pi \in \ASet$ and a fixed real number $\delta > 0$. Fix moreover an arbitrary initial point $(t,x,y)$ and define the control law $\pi_\delta$ by
\begin{align*}
\pi_{\delta}(s, x, y) = \begin{cases} \pi(s, x, y) &\mbox{for} \,\,(s, x, y) \in [t,t+\delta) \times \XSet \times \YSet\,,\\ \epi(s, x, y) &\mbox{for} \,\, (s, x, y) \in [t+\delta,T) \times \XSet \times \YSet\,. \end{cases}
\end{align*}
If, for all $\pi \in \ASet$, the following condition holds,
\begin{align*}
\underset{\delta\to 0}{\lim \inf} \quad \frac{J^{\hat\pi}(t,x,y) - J^{\pi_\delta}(t,x,y)}{\delta} \geq 0\,,
\end{align*}
then $\epi$ is referred to as an \textit{equilibrium control law}.
\end{definition}
For an equilibrium control law $\epi$, we define the equilibrium value function $\widehat{V}$ by
\[
\widehat{V}\left( t,x,y\right) :=J^{\epi }\left( t,x,y\right) \,.
\]

\section{Derivation of equilibrium controls} \label{sec:derivation_equilibrium}

This section develops the framework needed to characterize equilibrium investment policies. We begin with several preliminary definitions and then describe the dynamics of the state variables under the conditional measure that arises from our preference model. Building on these ingredients, we present a heuristic derivation of the eHJB governing equilibrium behavior, followed by a rigorous verification argument. We conclude the section by situating our eHJB within the broader literature, comparing its structure to existing formulations and highlighting key differences.

\subsection{Preliminary definitions and state process dynamics under conditional measures}\label{subsec:dynamics_under_different_measures}
Define, for any $\ybaro \in \YSet$,
\begin{equation}\label{eq:def_iota}
\varphi^{\ybaro}:= v\circ\left(u^{\gamma(\ybaro)}\right)^{-1}\,,
\end{equation}
which is twice continuously differentiable as the composition of two functions that are twice continuously differentiable. 

Let $f_{Y_T}(\ybaro; t, y)$ and $F_{Y_T}(\ybaro; t, y)$ denote, respectively, the conditional probability density function (PDF) and the conditional cumulative distribution function (CDF) of $Y_T$ given $Y_t = y$. Using this notation, we can rewrite the reward functional \eqref{eq:reward_functional} as
\begin{equation}
\label{eq:reward_functional_explicit}
\begin{split}
J^{\pi}(t, x, y) & = \Eop_{t, x, y}\left[ v \circ \left(u^{\gamma(Y_T)}\right)^{-1}\Big(\Eop_{t,x,y, \ybaro}\left[u^{\gamma(Y_T)}(X^{\pi}_T)\right]\Big)  \right]\, \\[0.2cm]
& =  \int_\mathcal{Y} \varphi^{\ybaro} \Big(\Eop_{t, x, y, \ybaro} \sBrackets{u^{\gamma(\ybaro)}\rBrackets{X^{\pi}_T}}\Big) f_{Y_T}(\ybaro; t, y)\, d\ybaro \\
& = \int_\mathcal{Y} \varphi^{\ybaro} \Big(\Eop_{t, x, y, \ybaro} \sBrackets{u^{\gamma(\ybaro)}\rBrackets{X^{\pi}_T}} \Big)\, dF_{Y_T}(\ybaro; t, y).
\end{split}
\end{equation}

The state process $\rBrackets{X^{\pi}, Y}$, conditional on $X^{\pi}_t = x$ and $Y_t = y$, satisfies the same SDEs \eqref{eq:SDE_X}-\eqref{eq:SDE_Y} with the initial condition $(x, y)$. In particular, under $\Pop_{t,x,y}$, we have:
 \begin{equation}
\begin{aligned}\label{eq:specific_SDE_X_Y_under_P}
             d X^\pi_t & = X^\pi_t (r+ \pi(t) (\mu_S-r))dt + X^\pi_t\pi(t) \sigma_S \rBrackets{ \rho d B^1_t + \sqrt{1 - \rho^2} d B^2_t},\\[0.2cm]
             dY_s & = \mu_Y(s, Y_s)ds  + \sigma_Y(s, Y_s) dB^1_s,
         \end{aligned}
     \end{equation}
     with $X_t = x$, $Y_t = y$, $\sigma_S > 0$, $\sigma_Y > 0$, and $dB^1_s dB^2_s = 0$ for every $s \in [t, T]$, i.e., $B^1$ and $B^2$ are still independent Brownian motions under $\Pop_{t,x,y}$.

Because the objective \eqref{eq:reward_functional_explicit} involves the conditional expectation $\Eop_{t,x,y,\ybaro}[\, \cdot \,]$, we need the dynamics of $\rBrackets{X^{\pi}, Y}$ under the conditional measure $\Pcond$. For this, we rely on a change-of-measure argument based on the transition density of $Y$; see \cite{DesmettreLeobacherRogers2021ChangeOfDrift} for the one-dimensional diffusion case. 

Let $p_Y(s, y; t,\ybaro)$  denote the transition density function of $Y$, i.e. the density of $Y_t = \ybaro$ given $Y_s = y$, for $0 \leq s \leq t \leq T$. The conditional density $f_{Y_T}$ is obtained as the special case $$f_{Y_T}(\ybaro; t, y) = p_Y(t, y; T, \ybaro).$$
It is convenient to keep both notations: we interpret $f_{Y_T}(\ybaro; t, y)$ as a function of $\ybaro \in \YSet$ for fixed  $(t, y) \in \TSet \times \YSet$, whereas $p_Y(s, y; t, \ybaro)$ will be treated as a function of $(s, y) \in \TSet \times \YSet$ for fixed $(t, \ybaro) \in [s, T] \times \YSet$. Denote by $\partial_y p_Y(s, y; T,\ybaro)$ the partial derivative of $p_Y(s, y; T,\ybaro)$ with respect to $y$. We then have the following result.
\begin{lemma}\label{lem:general_state_process_SDEs_under_Pcond}
Let $\Bbar^1$ and $\Bbar^2$ be two standard 
motions under $\Pop_{t,x,y, \ybaro}$. Then, under $\Pop_{t,x,y, \ybaro}$, for $s \in [t, T)$:
\begin{itemize}
\item The wealth process $X^\pi$ evolves as
\begin{equation}\label{eq:X_SDE_under_Pbar}
\begin{aligned}
dX^\pi_s & = X^\pi_s \rBrackets{r+ \pi(s) (\mu_S-r) + \pi(s)\sigma_S \rho \sigma_Y(s, Y_s) \partial_y \ln \big(p_Y(s, Y_s;{T},\ybaro)\big)}ds\\
& \quad + X^\pi_s\pi(s) \sigma_S \rBrackets{ \rho d \Bbar^1_s + \sqrt{1 - \rho^2} d \Bbar^2_s},\\
\end{aligned}
\end{equation}
with $X^\pi_t = x$ and $X^\pi_T = \lim_{t \to T} X^\pi_t$.
\item The process $Y$ evolves as
\begin{equation}\label{eq:Y_SDE_under_Pbar} 
\begin{aligned}
    d{Y}_s &= \rBrackets{\mu_Y(s, Y_s) + \sigma_Y^2(s, Y_s) \partial_y \ln \big(p_Y(s, {Y}_s;{T},\ybaro)}\big)ds + \sigma_Y\rBrackets{s, Y_s} d\Bbar^1_s,
\end{aligned}
\end{equation}
with $Y_t = y$ and $Y_T = \ybaro$.
\end{itemize}
\end{lemma}

\noindent \textit{Proof.} See Appendix \ref{proof:lemma_different_dynamics}. 

\medskip 
For later use, it is convenient to introduce compact notation for the drift and diffusion coefficients under the conditional measure $\Pop_{t,x,y,\ybaro}$:
\begin{align*}
\overline{\mu}_X(t,x,\pi)
&:= x\Bigl[r + \pi(\mu_S-r)
+ \pi\sigma_S \rho\, \sigma_Y(t,y)\,
\partial_y \ln p_Y(t,y;T,\ybaro)\Bigr],\\[0.2cm]
\overline{\sigma}_X(t,x,\pi)
&:= \left(\overline{\sigma}_{X,1}(t,x,\pi), \overline{\sigma}_{X,2}(t,x,\pi) \right) := 
\Bigl(
x\pi\sigma_S\rho,\;
x\pi\sigma_S\sqrt{1-\rho^2}
\Bigr),\\[0.2cm]
\overline{\mu}_Y(t,y)
&:= \mu_Y(t,y)
+ \sigma_Y^2(t,y)\,
\partial_y \ln p_Y(t,y;T,\ybaro),\\[0.2cm]
\overline{\sigma}_Y(t,y)
&:= \sigma_Y(t,y).
\end{align*}

\medskip

To express the eHJB in compact form, we now introduce the controlled differential operators associated with the dynamics under $\Pop_{t,x,y}$ and $\Pop_{t,x,y, \ybaro}$.
\begin{definition}
Let $X^{\pi}$ and $Y$ be given by \eqref{eq:SDE_X} and \eqref{eq:SDE_Y}, respectively, and let $\xi : \TSet \times \XSet \times \YSet \mapsto \mathbb{R}$ be a map such that $ \xi \in \textgoth{C}^{1,2, 2}\left(\TSet \times \XSet \times \YSet\right)$. 
(Given positive integers $p,q,r$, we write $\textgoth{C}^{p,q,r}(\mathbb{D})$ for the space of functions on the domain $\mathbb{D}$ that are continuously differentiable up to order $p,q,$ and $r$ in the respective arguments.)

For any admissible $\pi \in \ASet$, the controlled differential operator $\mathcal{D}^{\pi}$ under $\mathbb{P}_{t, x, y}$ is defined as follows:
\begin{equation*}\label{def:diff_operator_under_P}
    \begin{aligned}
        \mathcal{D}^{\pi}  \xi(t, x, y) & =  \partial_t \xi(t, x, y) + \mu_X(t,x,\pi(t, x, y)) \partial_x \xi(t, x, y) + \mu_Y(t,y) \partial_y \xi(t, x, y) \\
        &  \quad + \dfrac{1}{2} \norm{\sigma_X(t, x, y)}^2  \partial_{xx}\xi(t, x, y) + \dfrac{1}{2} \rBrackets{\sigma_Y(t, y)}^2 \partial_{yy}\xi(t, x, y)\\
        & \quad + \sigma_{X,1}(t, x, \pi(t, x, y)) \sigma_Y(t, y) \partial_{xy}\xi(t, x, y),
        \end{aligned}
\end{equation*}
where $\partial_{x}, \partial_y, \partial_{xx}, \partial_{xy}, \partial_{yy}$ denote the corresponding partial derivatives.

\medskip

Analogously, we define the controlled differential operator $\overline{\mathcal{D}}^{\pi}$ under $\Pop_{t,x,y, \ybaro}$ as follows:
\begin{equation*}\label{def:diff_operator_under_Pbar}
    \begin{aligned}
        \overline{\mathcal{D}}^{\pi}  \xi(t, x, y) & =  \partial_t \xi(t, x, y) + \overline{\mu}_X(t,x,\pi(t, x, y)) \partial_x \xi(t, x, y) + \overline{\mu}_Y(t,y) \partial_y \xi(t, x, y) \\
        &  \quad + \dfrac{1}{2} \norm{\overline{\sigma}_X(t, x, y)}^2  \partial_{xx}\xi(t, x, y) + \dfrac{1}{2} \rBrackets{\overline{\sigma}_Y(t, y)}^2 \partial_{yy}\xi(t, x, y)\\
        & \quad + \overline{\sigma}_{X,1}(t, x, \pi(t, x, y)) \overline{\sigma}_Y(t, y) \partial_{xy}\xi(t, x, y).
    \end{aligned}
\end{equation*}

For a constant control $\pi$, the operators are defined in the same way. 
\end{definition}

\subsection{Heuristic derivation of the eHJB}\label{subsec:heuristic_derivation_of_eHJB}
The reward functional \eqref{eq:reward_functional_explicit} does not fall directly within the general framework of \cite{BjoerkKhapkoMurgoci2021:TICT} (henceforth, BKM21), Section 15.5, whose most general objective (in our notation) when the dimensionality of the state process is $n=2$ has the form
\begin{equation}
\begin{aligned}\label{eq:reward_functional_general_BKM2021}
&\Eop_{t, x, y} \Bigg[ \int_t^T H(t, x, y, s,  X^{\pi}_s, Y_s, \pi(s,  X^{\pi}_s, Y_s)) \, ds\Bigg] \\[0.2cm]
& + \Eop_{t, x, y} \Big[ F\big(t, x, y, X^{\pi}_T, Y_T\big)\Big] + G\Big(t, x, y, \Eop_{t,x, y}[X^{\pi}_T], \Eop_{t,x, y}[Y_T]\Big) ,
\end{aligned}
\end{equation}
for possibly nonlinear functions $F, G, H$; cf. Eq. (15.13) therein. (In Section \ref{sec:relation_With_literature}, we return to this reward functional and discuss the corresponding eHJB.)

Two structural features of our preferences \eqref{eq:reward_functional_explicit} prevent a direct embedding into \eqref{eq:reward_functional_general_BKM2021}:
\begin{enumerate}
    \item the objective involves expectations conditional on a fixed  terminal state, $Y_T = \ybaro$, inside a nonlinear function of the expectation of the terminal state;
    \item the nonlinear function of the expectation of terminal state involves a continuum of terms $\varphi^{\ybaro} \rBrackets{\Eop_{t, x, y, \ybaro} \sBrackets{u^{\gamma(\ybaro)}\rBrackets{X^{\pi}_T}}}$, rather than one expectation of terminal state.
\end{enumerate}

In what follows, we then explain how to generalize \eqref{eq:reward_functional_general_BKM2021} to a form that includes \eqref{eq:reward_functional_explicit}. Feature (a) is addressed by using Lemma \ref{lem:general_state_process_SDEs_under_Pcond}, which describes the dynamics of $(X^{\pi}, Y)$ under $\Pop_{t,x,y, \ybaro}$. Feature (b), on the other hand, requires more work. Specifically, we need to approximate \eqref{eq:reward_functional_explicit} as a finite sum of different $G$-terms in \eqref{eq:reward_functional_general_BKM2021}, where we use the label \enquote{$G$-term} to refer to any nonlinear function of the expectation(s) of function(s) of the terminal value of the state process.

In the simplest case of a sum consisting of just one element, the approximating reward functional resembles \eqref{eq:reward_functional_general_BKM2021} with $F \equiv 0$, $H \equiv 0$ and one suitable $G\not \equiv 0$, with a minor difference that under the expectation operator there is a function of the terminal value of the state process (instead of just the terminal value of the state process itself). Therefore, for the heuristic derivation of the eHJB for our reward functional, we use the following reward functional as a starting point:
\begin{equation}
\label{eq:reward_functional_G_only_BKM2021}
   G\Big(t, x, y, \Eop_{t,x, y}[X^{\pi}_T], \Eop_{t,x, y}[Y_T]\Big). 
\end{equation}

As we explain in detail in Section \ref{sec:relation_With_literature}, the extended system characterizing an equilibrium control for \eqref{eq:reward_functional_G_only_BKM2021} involves an auxiliary function $g(t, x, y) = (g_1(t, x, y), g_2(t, x, y))$, where $g_1, g_2: \TSet \times \XSet \times \YSet \rightarrow \mathbb{R}$, and uses the following notation:
\begin{equation} \label{eq:notation_BKM21}
\begin{split}
(G \diamond g)(t, x, y) & := G(t, x, y, g_1(t, x, y), g_2(t, x, y)), \\[0.2cm]
\mathcal{H}^{\pi} g(t, x, y) & := G_{\xtil}(t, x, y, g_1(t, x, y), g_2(t, x, y)) \ \mathcal{D}^{\pi} g_1(t, x, y) \\
& \quad + G_{\ytil}(t, x, y, g_1(t, x, y), g_2(t, x, y)) \, \mathcal{D}^{\pi} g_2(t, x, y), \\
\mathcal{D}^{\widehat{\pi}} g(t, x, y) & := \rBrackets{\mathcal{D}^{\widehat{\pi}} g_1(t, x, y), \mathcal{D}^{\widehat{\pi}} g_2(t, x, y)}.
\end{split}
\end{equation}
The eHJB is then given by
\begin{align*}
0 & =\sup_{\pi \in \AMap(t,x,y)} \Big\{  \mathcal{D}^\pi V(t, x, y) - \mathcal{D}^{\pi}(G \diamond  g)(t, x, y) + \mathcal{H}^{\pi} g(t, x, y) \Big\},\\
(0, 0) & = \mathcal{D}^{\widehat{\pi}} g(t, x, y),  \\
V(T, x, y) & =  G(T, x, y, x, y),\\
g(T, x, y) &= (x, y),
\end{align*}
where $\widehat{\pi}$ denotes the control law that realizes the supremum in the first equation of the system; for more details, see Section \ref{sec:relation_With_literature}.

As stated in Section 16.3 of BKM21, it is possible to generalize the above eHJB to the case of 
\begin{equation*}
 G_1\Big(t, x,  y, \Eop_{t, x, y}[k_{1}^{1}(X^{\pi}_T)], \Eop_{t, x, y}[k_{1}^{2}(Y_T)]\Big)  , 
\end{equation*}
for some functions $k_{1}^{1}:\XSet \rightarrow \R$ and $k_{1}^{2}: \YSet \rightarrow \R$, where the superscript refers to the dimension of the state process $(X^{\pi}, Y)$ and the subscript refers to the number of expectations of a function of the terminal state process value.

For a one-dimensional state process $X^{\pi}$, another generalization is studied in \cite{KrygerNordfangSteffensen2020:MMRO}, where the eHJB is established for the case where the $G$-term depends on the conditional expectations of two different functions of the terminal state:
\begin{equation*}
    G_2\Big(t, x, \Eop_{t,x}[k_1(X^{\pi}_T)], \Eop_{t,x}[k_2(X^{\pi}_T)]\Big),
\end{equation*}
for $k_1, k_2:\XSet \rightarrow \R$.

In Remark 2 of the same paper, it is stated (yet not shown) that one can similarly obtain the eHJB for the case with $n>2$ different functions $k_i: \XSet \rightarrow \R$, $i = 1, \dots n$:
\begin{equation*}
    G_n\Big(t, x, \Eop_{t,x}[k_1(X^{\pi}_T)], \dots, \Eop_{t,x}[k_n(X^{\pi}_T)]\Big).
\end{equation*}
The function $G_n$ can be interpreted as an $n$-term finite aggregator of conditional expectations. Since finite sums provide natural discrete approximations of integrals, the expression above represents a tractable discretization of the continuum aggregation appearing in \eqref{eq:reward_functional_explicit}. In our problem, the integral averages certainty equivalents across the entire distribution of future preference states. By choosing the functions $k_i$ and the structure of $G_n$ appropriately, the sequence of discretized problems $(P_n)$ can approximate the original problem with continuous aggregation arbitrarily well. This observation allows us to construct the eHJB for our full objective by first analyzing the finite-dimensional case and then passing to the limit.

We proceed in two steps:
\begin{enumerate}
\item[1.] We construct a sequence of auxiliary problems $(P_n)$ that approximate the equilibrium investment problem associated with \eqref{eq:reward_functional_explicit}. Each problem $(P_n)$ replaces the integral aggregation over future preference states by a finite sum, enabling us to heuristically derive the eHJB in the finite-dimensional case $(P_n)$.
\item[2.] We then let $n \to \infty$ and interpret the integral in \eqref{eq:reward_functional_explicit} as the limit of these discrete approximations. In doing so, we obtain the eHJB for the original problem as the limiting case of the systems associated with $(P_n)$.
\end{enumerate}

\noindent \textit{Step 1: Sequence of approximating problems $(P_n)$}  

\medskip

\noindent Fix an arbitrary point $(t, x, y) \in [0, T) \times \XSet \times \YSet$. For any $n \in \N$, let $\mathcal{P}_{\YSet} := \cBrackets{\ybaro_0, \ybaro_1, \dots, \ybaro_n}$ be an arbitrarily chosen partition of $\YSet$, with $\ybaro_i \in \overline{\R}\,$ for every $i \in \cBrackets{0, 1, \dots,n}$, where $\overline{\R}:=\R \cup \cBrackets{-\infty, \infty}$. Define $\Delta F_i(t,y):=F_{Y_T}(\ybaro_{i+1}; t, y) - F_{Y_T}(\ybaro_{i}; t, y)$.
Then, we can approximate the integral as a finite sum as follows:
\begin{equation}
\label{eq:P_n_reward_functional_as_a_sum}
    \begin{aligned}
    J^{\pi}(t, x, y) & \approx \sum \limits_{i = 1}^{n} \varphi^{\ybaro_{i-1}} \Big(\Eop_{t, x, y, \ybaro_{i-1}} \sBrackets{u^{\gamma(\ybaro_{i - 1})}\rBrackets{X^{\pi}_T}}\Big) \Delta F_{i-1}(t,y) \\
    & =: G_{n}\Big(t, y, \Eop_{t, x, y, \ybaro_0} \sBrackets{u^{\gamma(\ybaro_0)}\rBrackets{X^{\pi}_T}},\dots,  \Eop_{t, x, y, \ybaro_{n - 1} } \sBrackets{u^{\gamma(\ybaro_{n - 1})}\rBrackets{X^{\pi}_T}}\Big) ,
    \end{aligned}
\end{equation}
where $G_n$ above does not have $x$ as its argument, though the derivation below can easily be extended to this case.

We define the $n$-th approximating problem $(P_n)$ as the one for which the reward functional is given by $$J^{\pi}_{n}(t, x, y):= G_{n}\Big(t, y, \Eop_{t, x, y, \ybaro_0} \sBrackets{u^{\gamma(\ybaro_0)}\rBrackets{X^{\pi}_T}},\dots,  \Eop_{t, x, y, \ybaro_{n - 1} } \sBrackets{u^{\gamma(\ybaro_{n - 1})}\rBrackets{X^{\pi}_T}}\Big).$$ Denote by $\epi_{n}$ an equilibrium investment strategy for $(P_n)$ and by $\widehat{V}_n(t, x, y)$ the respective equilibrium value function.

Let us derive the eHJB for the simplest case of $(P_{1})$. Choose an arbitrary partition $\mathcal{P}_{\YSet} = \cBrackets{\ybaro_0, \ybaro_1}$, i.e., $\ybaro_0$ is such that $F_{Y_T}(\ybaro_0; t, y) = 0$ and $\ybaro_1$ is such that $F_{Y_T}(\ybaro_1; t, y) = 1$. Then, $\Delta F_0(t, y) = 1$ and $(P_1)$ has the reward functional
\begin{equation}
\label{eq:P_1_reward_functional_compact}
\begin{split}
    J^{\pi}_1(t, x, y) & = \varphi^{\ybaro_0}\rBrackets{\Eop_{t, x, y, \ybaro_0}\sBrackets{u^{\gamma(\ybaro_0)}\rBrackets{X^{\pi}_T}}} \Delta F_0(t, y) \\
& = G_{1}\rBrackets{t, y, \Eop_{t, x, y, \ybaro_0}\sBrackets{u^{\gamma(\ybaro_0)}\rBrackets{X^{\pi}_T}}}. 
\end{split}
\end{equation}
Comparing \eqref{eq:P_1_reward_functional_compact} with \eqref{eq:reward_functional_G_only_BKM2021}, we observe that the two formulations are nearly identical, with only two conceptual differences. First, the appearance of the nonlinear function $u^{\gamma(\ybaro_0)}$ inside the expectation, which poses no structural difficulty; as noted in Section 16.3 of BKM21 and in \cite{KrygerNordfangSteffensen2020:MMRO}, such a modification can be incorporated simply by adjusting the terminal condition for the auxiliary function in the extended HJB system. Second, because the expectation in \eqref{eq:reward_functional_G_only_BKM2021} is also conditional on $Y_T = \ybaro$, the relevant state dynamics --and hence the differential operator appearing in the HJB-- must be taken under the conditional measure $\Pop_{t, x, y, \ybaro}$ rather than $\Pop_{t, x, y}$. Apart from this change of measure, the overall structure remains fully aligned with the framework of BKM21. (For completeness, we note that $G_1$ in \eqref{eq:P_1_reward_functional_compact} does not depend on $x$ and the conditional expectation of $Y_T$, in contrast to \eqref{eq:reward_functional_G_only_BKM2021}. If $G_1$ were to explicitly depend on these two objects, the arguments of this section could easily be adjusted to account for such dependence.)

Building on these observations, we now derive the eHJB for $(P_1)$. We begin by introducing the following notation:
\begin{align*}
    \rBrackets{G_1 \diamond g_1^{\ybaro_0}}(t,x,y) &:= G_1\big(t, y, g_1^{\ybaro_0}(t,x,y)\big) = \varphi^{\ybaro_0}\rBrackets{g_1^{\ybaro_0}(t, x, y)} \Delta F_0(t, y),\\[0.2cm]
    \overline{\mathcal{H}}^{\pi}_1 g_1^{\ybaro_0}(t, x, y) &:= \partial_{z_1} G_{1}\big(t, y, g_1^{\ybaro_0}(t, x, y)\big) \, \overline{\mathcal{D}}^{\pi} g_1^{\ybaro_0}(t, x, y), \\
    & = \rBrackets{\varphi^{\ybaro_0}}'\rBrackets{g_1^{\ybaro_0}(t, x, y)} \, \overline{\mathcal{D}}^{\pi} g_1^{\ybaro_0}(t, x, y) \, \Delta F_0(t, y).
\end{align*}
With this notation in place, the eHJB for $(P_1)$ takes the form
\begin{equation}
\label{eq:J_1:EHJB_PDE_1}
\begin{split}
0 & = \sup_{\pi \in \AMap(t,x, y)} \Big\{ \mathcal{D}^{\pi} V_1(t, x, y) - \mathcal{D}^{\pi} \rBrackets{G_1 \diamond g_1^{\ybaro_0}}(t,x,y) + \overline{\mathcal{H}}^{\pi}_1 g_1^{\ybaro_0}(t, x, y) \Big\} ,  \\[0.2cm]
0 & = \overline{\mathcal{D}}^{\epi_1} g_1^{\ybaro_0}(t, x, y), \\[0.2cm]
V_1(T, x, y) & = v(x), \\
g_1^{\ybaro_0}(T, x, y) & =  u^{\gamma(\ybaro_0)}(x),
\end{split}
\end{equation}
where $\epi_1$ realizes the supremum in the first equation of \eqref{eq:J_1:EHJB_PDE_1}. A slight modification of Theorem 15.2 in BKM21 verifies that, under certain regularity conditions, $\epi_1$ solving the above extended system is indeed an equilibrium control for $(P_1)$.

\medskip

Applying similar heuristic reasoning as in Section 15.3.1 of BKM21 to the problem $(P_n)$, whose reward functional is given by \eqref{eq:P_n_reward_functional_as_a_sum}, we obtain the natural extension of the system derived for $(P_1)$. To express this compactly, we introduce the notation
\begin{align}
    \rBrackets{G_n\diamond \rBrackets{g_1^{\ybaro_0}, \dots,  g_n^{\ybaro_{n-1}}}} \rBrackets{t, x, y} & := G_n\rBrackets{t, y, g^{\ybaro_0}_1\rBrackets{t, x, y}, \dots, g^{\ybaro_{n-1}}_n\rBrackets{t, x, y}} \notag \\
    & = \sum \limits_{i = 1}^{n} \varphi^{\ybaro_{i - 1}}\rBrackets{g_i^{\ybaro_{i - 1}}(t, x, y))}  \Delta F_{i - 1}(t,y), \notag  \\
    \overline{\mathcal{H}}^{\pi}_n \rBrackets{\rBrackets{g_i^{\ybaro_{i-1}}}_{i=1,\dots,n}}(t,x,y) & := \sum \limits_{i = 1}^{n } \rBrackets{\varphi^{\ybaro_{i-1}}}'\rBrackets{ g_i^{\ybaro_{i - 1}}(t, x, y))} \notag \\
    &  \qquad \times \overline{\mathcal{D}}^{\pi} g_i^{\ybaro_{i -  1}}(t, x, y) \, \Delta F_{i - 1}(t, y). \notag 
\end{align}
These expressions parallel exactly the one-component case, except that each discrete preference scenario contributes its own auxiliary function $g_i^{\ybaro_{i-1}}$ and sensitivity term weighted by the corresponding probability mass $\Delta F_{i-1}(t,y)$.

With this notation in place, the eHJB for $(P_n)$ takes the form
\begin{equation}
\label{eq:J_n:EHJB_PDE_1}
\begin{aligned}
0 & = \sup_{\pi \in \AMap(t,x, y)} \Bigg\{ \mathcal{D}^{\pi} V_n(t, x, y) - \mathcal{D}^{\pi}\rBrackets{G_n\diamond \rBrackets{g_1^{\ybaro_0}, \dots,  g_n^{\ybaro_{n-1}}}} \rBrackets{t, x, y}  \Bigg. \\
& \qquad \qquad \qquad \Bigg. + \overline{\mathcal{H}}^{\pi}_n \rBrackets{\rBrackets{g_i^{\ybaro_{i-1}}}_{i=1,\dots,n}}(t,x,y) \Bigg\},  \\
0 & = \overline{\mathcal{D}}^{\epi_n} g_i^{\ybaro_{i - 1}}(t, x, y), \qquad i \in \cBrackets{1,\dots, n},\\
V_n(T, x, y) &  = v(x), \\
g_i^{\ybaro_{i - 1}}(T, x, y) & =  u^{\gamma(\ybaro_{i - 1})}(x), \, \qquad i \in \cBrackets{1,\dots, n},
\end{aligned}
\end{equation}
where $\epi_n$ realizes the supremum in the first equation of \eqref{eq:J_n:EHJB_PDE_1}. Once more, an appropriate adaptation of the verification argument in BKM21 shows that, under suitable smoothness and integrability conditions, the control $\epi_n$ obtained from this system constitutes an equilibrium control for $(P_n)$.

\newpage
\noindent \textit{Step 2: eHJB for the limiting case} 

\medskip

To obtain the eHJB associated with the original objective functional \eqref{eq:reward_functional_explicit}, we now pass from the discrete problems $(P_n)$ to their continuous counterpart. This motivates introducing the limiting objects
\begin{align}
    G_{\infty}(t,x,y)&:=\lim_{n \to \infty} \rBrackets{G_n \diamond \rBrackets{g_1^{\ybaro_0}, \dots,  g_n^{\ybaro_{n-1}}}} \rBrackets{t, x, y} \notag \\
    & =   \int_\mathcal{Y} \varphi^{\ybaro}\rBrackets{\gybar{t, x, y}} dF_{Y_T}(\ybaro; t, y) \label{eq:G_infty}
    , \\
    \overline{\mathcal{H}}^{\pi}\rBrackets{\rBrackets{g^{\ybaro}}_{\ybaro \in \YSet}}(t, x, y) &:= \lim_{n \to \infty} \overline{\mathcal{H}}^{\pi}_n \rBrackets{\rBrackets{g_i^{\ybaro_{i-1}}}_{i=1,\dots,n}}(t,x,y) \notag \\
    & = \int_\mathcal{Y} \rBrackets{\varphi^{\ybaro}}'\rBrackets{\gybar{t, x, y}}  \, \overline{\mathcal{D}}^{\pi} \gybar{t, x, y} \, dF_{Y_T}(\ybaro; t, y). \label{eq:barH_pi}
\end{align}

We are now equipped to write down the eHJB characterizing an equilibrium control for the original, infinite-dimensional aggregation problem.

\paragraph{The extended HJB system.}  The system consists of the following coupled relations:
\begin{align}
0 & = \sup_{\pi \in \AMap(t,x, y)} \Bigg\{ \mathcal{D}^{\pi} V(t, x, y) - \mathcal{D}^{\pi} G_{\infty}(t,x,y) \label{eq:J_inf:EHJB_PDE_1}   + \overline{\mathcal{H}}^{\pi}\rBrackets{\rBrackets{g^{\ybaro}}_{\ybaro \in \YSet}}(t, x, y)  \Bigg\} , \tag{$S1$} \\
0 & = \overline{\mathcal{D}}^{\epi} \gybar{t, x, y}, \qquad \bar{y} \in \YSet, \label{eq:J_inf:EHJB_PDE_2} \tag{$S2$} \\
v(x)  & = V(T, x, y) \label{eq:J_inf:terminal_condition_V} \tag{$S3$} \\
 u^{\gamma(\ybaro)}(x)  & = \gybar{T, x, y}, \qquad \bar{y} \in \YSet,  \label{eq:J_inf:terminal_condition_g} \tag{$S4$}
\end{align}
where $\epi$ realizes the supremum in \eqref{eq:J_inf:EHJB_PDE_1}.

\medskip
The following section provides a verification theorem establishing that a solution of \eqref{eq:J_inf:EHJB_PDE_1}-\eqref{eq:J_inf:terminal_condition_g} indeed yields an equilibrium strategy.

\subsection{Verification results}\label{subsec:verification_results}

For the proof of the main theorem, we will require suitable integrability conditions. These are outlined in the following definition of an $\mathcal{L}^2$ function space.

\begin{definition}
\label{def:admissible_space_of_functions}
Fix an arbitrary control $\pi \in \ASet$. A function $\xi:\TSet \times \XSet \times \YSet \rightarrow \mathbb{R}$ is said to belong to the space $\mathcal{L}^2(X^{\pi}, Y)$ if, for any $(t, x, y) \in [0, T) \times \XSet \times \YSet$, there exists a constant $\bar{\delta} \in (0, T - t)$ such that
\begin{align*}
& \mathbb{E}_{t, x, y} \Bigg[ \sup_{0 \leq \delta \leq \bar{\delta}} \Bigg\vert \int_{t}^{t+\delta}\dfrac{1}{\delta}\mathcal{D}^{\pi}\xi(s, X^{\pi}_s, Y_s) ds \; \Bigg\vert \Bigg.\\
& \hspace{2cm} \Bigg. + \int_{t}^{t+\bar{\delta}} \norm{\partial_{x} \xi(s, X^{\pi}_s, Y_s)\sigma_X\rBrackets{s,  X^{\pi}_s, \pi(s)}}^{2} ds \Bigg.\\
& \hspace{2cm} \Bigg. + \int_{t}^{t+\bar{\delta}}\Big( \partial_{y} \xi(s, X^{\pi}_s, Y_s) \sigma_Y\rBrackets{s, Y_s} \Big)^{2} ds \Bigg] < \infty.
\end{align*}
Analogously, we say that $\xi:\TSet \times \XSet \times \YSet \rightarrow \mathbb{R}$ belongs to the space $\overline{\mathcal{L}}^2(X^{\pi}, Y)$ if, for any $(t, x, y) \in [0, T) \times \XSet \times \YSet$, there exists a constant $\bar{\delta} \in (0, T - t)$ such that
\begin{align*}
& \mathbb{E}_{t, x, y} \Bigg[ \sup_{0 \leq \delta \leq \bar{\delta}} \Bigg\vert \int_{t}^{t+\delta}\dfrac{1}{\delta}\overline{\mathcal{D}}^{\pi}\xi(s, X^{\pi}_s, Y_s) ds \; \Bigg\vert \Bigg.\\
& \hspace{2cm} \Bigg. + \int_{t}^{t+\bar{\delta}} \norm{\partial_{x} \xi(s, X^{\pi}_s, Y_s)\overline{\sigma}_X\rBrackets{s,  X^{\pi}_s, \pi(s)}}^{2} ds \Bigg.\\
& \hspace{2cm} \Bigg. + \int_{t}^{t+\bar{\delta}}\Big( \partial_{y} \xi(s, X^{\pi}_s, Y_s) \overline{\sigma}_Y\rBrackets{s, Y_s} \Big)^{2} ds \Bigg] < \infty.
\end{align*}
\end{definition}

\medskip

We now introduce a family of auxiliary functions parameterized by $\ybaro \in \YSet$:  
\begin{align}
g^{\pi;\ybaro}(t,x, y) = \Eop_{t,x,y, \ybaro}\left[ u^{\gamma(Y_T)}(X^{\pi}_T)\right]  \,,
\end{align}
where, by construction, $y$ and $\ybaro$ must coincide at $t = T$. 

In the following lemma, we derive a recursive representation for each function in $\rBrackets{g^{\pi;\ybaro}}_{\ybaro \in \YSet}$ and also characterize it via a PDE. Its proof follows the same steps as Lemma 3.5 in \cite{Lindensjoe2019:ORL} or Lemma 3.7 in \cite{DeGennaroAquino2024equilibrium}, with the only difference that we work here under the conditional measure $\Pop_{t, x, y, \ybaro}$ instead of $\Pop_{t, x, y}$.
\begin{lemma}\label{lem:auxiliary_function_characterization}
For any admissible control $\pi \in \ASet$ and $\ybaro \in \YSet$, for any $\delta \in (0, T - t]$, the function $g^{\pi;\ybaro}(t, x, y)$ satisfies the recursive relation
\begin{equation*}
g^{\pi;\ybaro}(t, x, y) = \Eop_{t,x,y, \ybaro}\sBrackets{g^{\pi;\ybaro}\rBrackets{t + \delta, X^{\pi}(t+\delta), Y(t + \delta)}},
\end{equation*}
and the terminal condition
\begin{equation*}
g^{\pi;\ybaro}(T, x, y) = u^{\gamma(\ybaro)}(x).
\end{equation*}

\medskip
\noindent Moreover, if $g^{\pi;\ybaro} \in \textgoth{C}^{1, 2, 1}\left(\TSet \times \R \times \R \right) \cap \overline{\mathcal{L}}^2\rBrackets{X^{\pi}, Y}$, then, for $t \in [0,T)$, $g^{\pi;\ybaro}(t, x, y)$ satisfies the PDE
\begin{equation*}
\overline{\mathcal{D}}^{\pi} g^{\pi;\ybaro}(t, x, y) = 0.
\end{equation*}
\end{lemma} 

We are now in a position to state our main result.
\begin{theorem}[Verification theorem]\label{th:verification}
Assume that the functions $V(t, x, y)$, $G_{\infty}(t, x, y)$, and the family $\rBrackets{\gybar{t, x, y}}_{\ybaro \in \YSet}$ satisfy the following properties:
\begin{enumerate}
\item[(C1)] The arg sup in \eqref{eq:J_inf:EHJB_PDE_1} exists, is denoted by $\epi$, and it is an admissible control.
\item[(C2)] The triplet $\left(V, G_{\infty}, \rBrackets{g^{\ybaro}}_{\ybaro \in \YSet}\right)$ solves the system \eqref{eq:J_inf:EHJB_PDE_1}-\eqref{eq:J_inf:terminal_condition_g}.
\item[(C3)] For every $\ybaro \in \YSet$, $$g^{\ybaro} \in \mathcal{C}^{1,2, 2}\left(\TSet \times \XSet \times \YSet\right)$$  
and
$$V \in \mathcal{C}^{1,2, 2}\left(\TSet \times \XSet \times \YSet\right), G_{\infty} \in \mathcal{C}^{1,2, 2}\left(\TSet \times \XSet \times \YSet\right).$$
\item[(C4)] For every $\ybaro \in \YSet$ and $\pi \in \ASet$, $$g^{\ybaro} \in \overline{\mathcal{L}}^2\rBrackets{X^{\pi}, Y^{\pi}}, $$
and for every $\pi \in \ASet$,
$$V \in \mathcal{L}^2\rBrackets{X^{\pi}, Y^{\pi}}, G_{\infty} \in \mathcal{L}^2\rBrackets{X^{\pi}, Y^{\pi}}.$$
\end{enumerate}
Then:
\begin{enumerate}
\item[(R1)] $g^{\ybaro}(t, x, y) = g^{\epi;\ybaro}(t, x, y)$, and admits the probabilistic representation
\begin{align*}
g^{\ybaro}(t, x, y) &= \Eop\sBrackets{u^{\gamma(Y_T)}\rBrackets{X^{\epi}_T} \vert \, X^{\epi}_t = x, Y_t = y, Y_T = \ybaro} \\[0.2cm]
& = \Eop\sBrackets{u^{\gamma(\ybaro)}\rBrackets{X^{\epi}_T}\vert \, X^{\epi}_t = x, Y_t = y}.
\end{align*}
\item[(R2)] For $\epi$ realizing the sup in \eqref{eq:J_inf:EHJB_PDE_1}, the objective can be written as $$J^{\epi}(t, x, y) = \int_\mathcal{Y} \varphi^{\ybaro} \rBrackets{g^{\ybaro}(t, x, y)}\,dF_{Y_T}(\ybaro; t, y).$$
\item[(R3)] $\epi$ is an equilibrium investment strategy.
\item[(R4)] The equilibrium value function is given by $$\widehat{V}(t, x, y) = V(t, x, y)  = \int_\mathcal{Y} \varphi^{\ybaro} \rBrackets{g^{\ybaro}(t, x, y)} \, dF_{Y_T}(\ybaro; t, y),$$
and admits the probabilistic representation \eqref{eq:reward_functional_explicit} for $\pi = \epi$.
\end{enumerate}
\end{theorem}

\noindent \textit{Proof.} See Appendix \ref{proof_verification_theorem}. 

\medskip

The verification theorem allows us to simplify the eHJB \eqref{eq:J_inf:EHJB_PDE_1}-\eqref{eq:J_inf:terminal_condition_g} in a way analogous to what is done in Section 16.1 of BKM21. From (R4) and \eqref{eq:G_infty}, we in fact notice that 
\begin{equation*}
V(t, x, y)  = G_{\infty}(t,x, y).
\end{equation*}
Therefore, the first two terms under the supremum in \eqref{eq:J_inf:EHJB_PDE_1} cancel out.

\begin{corollary} 
\label{cor:rewriting_HJB_system}
Using \eqref{eq:barH_pi}, the eHJB \eqref{eq:J_inf:EHJB_PDE_1}-\eqref{eq:J_inf:terminal_condition_g}  can be written in the following equivalent form:
\begin{align}
0 & = \sup \limits_{\pi \in \AMap(t, x, y)} \overline{\mathcal{H}}^{\pi}\rBrackets{\rBrackets{g^{\ybaro}}_{\ybaro \in \YSet}}(t, x, y), \tag{$\overline{S1}$} \label{EHJB_eq_1} \\
0 & = \overline{\mathcal{D}}^{\epi} \gybar{t, x, y}, \qquad  \, \ybaro \in \YSet, \tag{$\overline{S2}$} \label{EHJB_eq_2}\\
u^{\gamma(\ybaro)}(x) &= \gybar{T, x, y}, \qquad \, \ybaro \in \YSet, \tag{$\overline{S3}$} \label{EHJB_eq_3}
\end{align}
where $\epi$ realizes the supremum in \eqref{EHJB_eq_1}.
\end{corollary}

\subsection{Relation to existing formulations}
\label{sec:relation_With_literature}

To place our eHJB in context, we relate it to the general framework of time-inconsistent control developed by BKM21. We begin by recalling the eHJB associated with the objective function \eqref{eq:reward_functional_general_BKM2021}, adapted to the case when the state space is two-dimensional. For clarity, we consistently translate their notation into ours.

For functions $g = (g_1, g_2)$ and $G$, recall the definitions used by BKM21 introduced in \eqref{eq:notation_BKM21}. In addition, for a function $f$, they write
\begin{equation*}
f^{s \xtil \ytil}(t, x, y) := f(t, x, y, s, \xtil, \ytil), \quad \mbox{with }  s, \xtil, \ytil \; \mbox{ seen as fixed values.}
\end{equation*}
With these definitions, the full characterization of the value function $V$, the family of auxiliary value functions $f^{s \xtil \ytil}(t, x, y)$, and the function $g$, takes the form:
\begin{equation}
\label{eq:bigsystem_BKM}
\begin{split}
0  & = \sup_{\pi \in \AMap(t,x,y)} \Big\{ \mathcal{D}^\pi V(t, x, y)+ H(t, x, y, t, x, y, \pi) - \mathcal{D}^\pi f(t, x, y, t, x, y), \\
& \hspace{1.5cm} + \mathcal{D}^\pi f^{txy}(t, x, y) - \mathcal{D}^{\pi}(G \diamond  g)(t, x, y) + (\mathcal{H}^{\pi} g)(t, x, y) \Big\},  \\
0 & =\mathcal{D}^{\widehat{\pi}} f^{s \xtil \ytil}(t, x, y) + H(s, \xtil, \ytil, t, x, y, \widehat{\pi}), \quad (s, \tilde{x}, \tilde{y}) \in [0, T) \times \XSet \times \YSet, \\
(0, 0) &= \mathcal{D}^{\widehat{\pi}} g(t, x, y),  \\
V(T, x, y) & = F(T, x, y, x, y) + G(T, x, y, x, y), \\
f^{s \xtil \ytil}(T, x, y) &= F(s, \xtil, \ytil, x, y), \quad (s, \tilde{x}, \tilde{y}) \in [0, T) \times \XSet \times \YSet, \\
g(T, x, y) &=  (x, y),
\end{split}
\end{equation}
where $\widehat{\pi}$ realizes the supremum in the first equation of \eqref{eq:bigsystem_BKM}.

To understand how our system relates to the general framework presented above, we need to examine what happens when the general reward functional \eqref{eq:reward_functional_general_BKM2021}  contains only the nonlinear $G$-term corresponding to \eqref{eq:reward_functional_G_only_BKM2021}. This case arises by setting $H \equiv 0$ and $F\equiv 0$.

Under this restriction, several simplifications occur in the system \eqref{eq:bigsystem_BKM}. Firstly, the second and fifth equations are redundant. Secondly, by noticing that $V = G \diamond g$, the two terms $(\mathcal{D}^\pi V)(t, x, y)$ and $\mathcal{D}^{\pi}(G \diamond  g)(t, x, y)$ become identical. This gives the system
\begin{equation}
\begin{aligned}
0 & = \sup_{\pi \in \AMap(t,x,y)}  (\mathcal{H}^{\pi} g)(t, x, y), \\
(0, 0) & = \mathcal{D}^{\widehat{\pi}} g(t, x, y), \\
g(T, x, y) & =  (x, y),
\end{aligned}
\end{equation}
which is structurally similar to \eqref{EHJB_eq_1}--\eqref{EHJB_eq_3}.

\bigskip

As for the extended system in \cite{DesmettreSteffensen2023:MF}, Theorem 3.5, observe that $U_t(t,x)$ and all the terms in the first two lines under the infimum in their Eq. (10) cancel out as a consequence of their Eq. (20). After these cancellations, their extended system can be rewritten in the following compact form:
\begin{equation}
\label{eq:eHJB_DesmettreSteffensen2023}
\begin{aligned}
0 & = \sup_{\pi} \int \iota^{\gamma}(Y^\gamma(t,x)) \, \mathcal{D}^{\pi}Y^{\gamma}(t,x) \, \mathrm{d}\Gamma(\gamma),   \\
0 & = \mathcal{D}^{\epi}Y^{\gamma}(t,x), \qquad \text{for each realization of } \gamma,\\
u^{\gamma}(x) & = Y^\gamma(T,x), \qquad \text{for each realization of } \gamma,
\end{aligned}
\end{equation}
which is identical to our eHJB for $v(x) = x$ once we translate their notation into ours:
\begin{equation*}
    \begin{split}
         & \gamma \rightarrow \gamma(\ybaro), \quad \iota^\gamma \rightarrow \rBrackets{\varphi^{\ybaro}}',\quad  Y^\gamma(t,x) \rightarrow g^{\ybaro}(t,x,y), \\
        & u^{\gamma} \rightarrow u^{\gamma(\ybaro)},\quad d\Gamma(\gamma) \rightarrow dF_{Y_T}(\ybaro;t,y).
    \end{split}
\end{equation*}

\bigskip
Finally, we relate our eHJB to the system in Eq. (9) of \cite{ChenGuanLiang2025}, reported below:
\begin{equation}
\begin{split}
0 & =\sup_{\pi \in \Pi} \left\{ \sum_{k \in \mathcal{S}}  \left( (u^k)^{-1} \right)' \left( f^{i,k}(t,x) \right) \;  \mathscr{A}^\pi f^{i,k}(t,x) \; p(t,i,k) \right\} = 0, \label{eq:CGL2025_HJB_sytem_eq_1} \\
0 & =\mathscr{A}^\pi f^{i,j}(t,x) = 0, \qquad j \in \mathcal{S}, \\
f^{i,j}(T,x) & = u^j(x), \qquad j \in \mathcal{S}. 
\end{split}
\end{equation}
Here $\mathcal{S} :=\cBrackets{1, \dots, n}$, with $n \in \mathbb{N}$, is the state space of the Markov chain $\epsilon :=\rBrackets{\epsilon_t}_{\tin}$ driving both risk aversion and  market coefficients, $p(t, i, k) = \Pop\rBrackets{\epsilon_T = k \, \vert \ \epsilon_t = i}$, for every $\tin$ and $i,k \in \mathcal{S}$, $\Pi$ is the set of admissible strategies, and $\mathscr{A}^{\pi}$ is the controlled infinitesimal generator of the regime-switching diffusion.  Naturally, \eqref{eq:CGL2025_HJB_sytem_eq_1} corresponds to the special case of our \eqref{EHJB_eq_1} under $v(x) = x, \rho = 0$, and the dictionary 
\begin{equation*}
    \begin{split}
        & \Sigma \rightarrow \int, \quad \mathcal{S} \rightarrow \YSet, \quad k \rightarrow \ybaro, \quad   j \rightarrow \ybaro, \quad i \rightarrow y, \quad \rBrackets{u^k}^{-1} \rightarrow \varphi^{\ybaro},   \\
        & p(t, i, k) \rightarrow dF_{Y_T}(\ybaro; t, y), \quad f^{i,j}(t,x) \rightarrow g^{\ybaro}(t, x, y). 
    \end{split}
\end{equation*}
Due to the assumption of independence between the Markov chain $\epsilon$ and the Brownian motion driving the wealth process in \cite{ChenGuanLiang2025}, conditioning on $\epsilon_T = k$, for some $k \in \mathcal{S}$, does not change the law of the wealth process. In contrast, our setting allows for arbitrary correlation between $X$ and $Y$ via $\rho \in [-1, 1]$. Conditioning on $Y_T = \ybaro$ therefore changes the law of $X$, which requires a change of the differential operator from $\mathcal{D}$ (under $\Pop_{t, x, y}$) to $\overline{\mathcal{D}}$ (under $\Pop_{t, x, y, \ybaro}$).

\bigskip

\section{Application: state-dependent CRRA utility}\label{sec:application}

In this section, we apply the general equilibrium framework developed in Sections \ref{sec:problem_formulation} and \ref{sec:derivation_equilibrium} to a tractable case in which the investor's relative risk aversion is driven by an exogenous factor that follows an arithmetic Brownian motion. 

First, we describe the preference specification and derive the specialized form of the eHJB. Second, we analyze the resulting equilibrium policy numerically and provide some intuition on the underlying economic mechanisms.

\subsection{Preference specification and equilibrium investment}

We consider preferences that are CRRA with a state-dependent relative risk aversion. The intertemporal variation in risk attitudes is driven by an exogenous factor, denoted by $Y$, which evolves according to the arithmetic Brownian motion
\begin{equation}\label{eq:SDE_Y_ABM}
dY_t = \mu_Y dt + \sigma_Y dB^1_t,
\end{equation}
for constants $\mu_Y\in\mathbb{R}$ and $\sigma_Y>0$, with slight abuse of notation compared to \eqref{eq:SDE_Y}. 

For a fixed realization of $Y_{T} = \ybaro$, we model the risk aversion coefficient by
\begin{equation} \label{eq:gamma_fun}
\gamma(\ybaro) := e^{\ybaro},
\end{equation}
which induces the CRRA utility
\begin{equation} \label{eq:CRRA_spec}
    u^{\gamma(\ybaro)}(x) = \dfrac{x^{1-\gamma(\ybaro)}}{1-\gamma(\ybaro)}.
\end{equation}
Thus, the investor behaves as a standard CRRA agent along the wealth dimension, while changes in $Y$ alter her effective risk aversion over time.

The corresponding evaluation of a portfolio strategy $\pi$ is recalled here for convenience:
\begin{equation*}
J^\pi(t,x,y)
=
\int_{\mathcal{Y}}
\varphi^{\ybaro}\!\left(
\mathbb{E}_{t,x,y,\bar y}\big[u^{\gamma(\bar y)}(X^\pi_T)\big]
\right)
f_{Y_T}(\bar y;t,y)\,d\bar y.
\end{equation*}
In the general framework of Section \ref{sec:derivation_equilibrium}, $v$ is an arbitrary increasing, concave, and differentiable function. Here, we adopt the logarithmic form
\begin{equation} \label{eq:log_aggregator}
    v(x) = \ln(x),
\end{equation}
which is helpful for tractability reasons as it naturally aligns with the multiplicative structure of CRRA preferences. (This form of aggregator for certainty equivalents is also used in \cite{BS21}, albeit in a different context.)

In addition, recall that the density $f_{Y_T}(\bar y;t,y)$ is the PDF of $Y_T$ conditional on $Y_t = y$, which is Gaussian, since \eqref{eq:SDE_Y_ABM} implies
\begin{equation*}\label{eq:ABM_Y_T_given_Y_t}
    Y_T \, \vert \, Y_t = y \sim \mathcal{N}\left( y + \mu_Y(T-t), \sigma_Y^2(T-t) \right).
\end{equation*}

\medskip

To construct the equilibrium policy, we next adapt the general eHJB from Section \ref{sec:derivation_equilibrium} to this CRRA setting. The first step is to identify the dynamics of the state process $(X^{\pi}, Y)$ under the family of conditional measures appearing in the equilibrium equations. The next result follows from Lemma~\ref{lem:general_state_process_SDEs_under_Pcond} upon inserting the diffusion specification for $Y$.
\begin{corollary}\label{cor:special_case:state_process_SDEs_under_Pcond}
Let $\Bbar^1$ and $\Bbar^2$ be two standard Brownian motions under the conditional measure $\Pop_{t,x,y, \ybaro}$. Then, for $s \in [t, T)$:
\begin{itemize}
    \item The preference factor $Y$ satisfies
\begin{equation*}
d Y_s = \frac{\ybaro - Y_s}{T - s}ds + \sigma_Y d \Bbar^1_s,
\end{equation*}
with $Y_t = y$ and $Y_T = \ybaro$.
\item The wealth process $X^{\pi}$ satisfies
\begin{equation*}
\begin{aligned}
dX^\pi_s & = X^\pi_s \rBrackets{r+ \pi(s) (\mu_S-r) + \pi(s)\rho \frac{\sigma_S }{\sigma_Y} \frac{\ybaro - Y_s - \mu_Y(T - s)}{T - s}}ds\\
& \qquad + X^\pi_s\pi(s) \sigma_S \rBrackets{ \rho d \Bbar^1_s + \sqrt{1 - \rho^2} d \Bbar^2_s},
\end{aligned}
\end{equation*}
with $X^\pi_t = x$ and $X^\pi_T = \lim_{t \to T} X^\pi_t$.
\end{itemize}
\end{corollary}

\noindent \textit{Proof.} See Appendix \ref{proof_corollary1_application}.

\medskip

The process $Y$ thus becomes an arithmetic Brownian bridge under $\Pop_{t,x,y, \ybaro}$, whereas $X^{\pi}$ carries an additional drift adjustment reflecting both the conditioning on $Y_T = \bar{y}$ and the correlation between the underlying shocks.

\medskip

The following proposition provides the semi-explicit representation of an equilibrium policy. Its expression depends on a family of auxiliary functions $h^{\bar{y}}(t,y)$ which solve a coupled nonlinear PIDE.

\begin{proposition}\label{prop:ln-CRRA_preferences}
For a reward functional of the form in \eqref{eq:reward_functional}, consider a CRRA specification \eqref{eq:gamma_fun}-\eqref{eq:CRRA_spec} with logarithmic certainty equivalent aggregator \eqref{eq:log_aggregator}. In this setting, the equilibrium investment policy depends only on $(t,y)$ and is given by
\begin{equation} \label{eq:equilibrium_policy_semiexplicit_main}
\begin{split}
     \widehat{\pi}(t,y) & = \dfrac{\mu_S - r + \rho \sigma_S \sigma_{Y} \displaystyle{\int_\mathcal{Y} \dfrac{\partial_y h^{\bar{y}}(t,y) }{h^{\bar{y}}(t,y)}f_{Y_T}(\bar{y}; t,y)d\bar{y}}}{\sigma_S^2 \exp\left(y + \mu_Y(T-t) + \dfrac{1}{2}\sigma_{Y}^{2}(T-t)\right)},
  \end{split}  
\end{equation}  
where, for each $\bar{y} \in \YSet$, the function $ h^{\bar{y}}$ solves
\begin{equation} \label{eq:PIDE_finalform_main}
    \begin{split}
 0 & = \partial_t h^{\bar{y}}(t,y) + \left( \dfrac{\bar{y}-y}{T-t} + \rho \widehat{\pi} \sigma_S \sigma_{Y}  \big(1-e^{\bar{y}}\big) \right) \partial_y h^{\bar{y}}(t,y) + \dfrac{1}{2}\sigma_{Y}^2 \partial_{yy} h^{\bar{y}}(t,y) \\
& +\left( r+ \widehat{\pi}\left(\mu_S-r + \rho \dfrac{\sigma_S}{\sigma_{Y}} \left( \dfrac{\bar{y}-y}{T-t}-\mu_Y \right) \right)   - \dfrac{1}{2} \widehat{\pi}^2 \sigma_S^2  \, e^{\bar{y}} \right) \big(1-e^{\bar{y}}\big)  h^{\bar{y}}(t,y), \\
h^{\bar{y}}(T,y) & = 1.
    \end{split}
\end{equation}
\end{proposition}

\noindent \textit{Proof.} See Appendix \ref{proof_corollary2_application}.

\medskip

The relation defining the equilibrium control in \eqref{eq:equilibrium_policy_semiexplicit_main} and the PIDE for $h^{\bar{y}}$ in \eqref{eq:PIDE_finalform_main} are jointly nonlinear and fully coupled, except in the special case $\rho = 0$. The numerical computations reported below are based on solving this forward-backward system.  

\subsection{Interpretation and numerical analysis}

Let us comment on the equilibrium investment policy in \eqref{eq:equilibrium_policy_semiexplicit_main}. Since the exponential factor appearing in the denominator equals the conditional expectation of $\gamma(Y_T)$ given $Y_t = y$, we may rewrite $\epi(t,y)$ as 
   \begin{equation*}
    \epi(t,y) =  \dfrac{\mu_S - r}{\sigma_S^2 \mathbb{E}_{t,y}\left[\gamma(Y_T)\right]} + \dfrac{\rho \sigma_Y }{\sigma_S \mathbb{E}_{t,y}\left[\gamma(Y_T)\right]}
\int_\mathcal{Y} \frac{\partial_y h^{\bar{y}}(t,y)}{h^{\bar{y}}(t,y)} f_{Y_T}(\bar{y};t,y) \,d\bar{y}. 
\end{equation*}  
The first term is the familiar myopic (Merton-type) demand, but with the constant risk aversion coefficient replaced by the conditional expectation of its terminal realization. The second term is what we call a \textit{preference-hedging demand}, 
for it captures the agent's intention to adjust today's exposure in anticipation of how her future risk aversion may evolve.  Crucially, this hedging term depends not only on preference dynamics but also on how these dynamics interact with the market. When $\rho \neq 0$, preference shocks and asset return shocks are partially correlated, giving the investor a channel to hedge the stochastic evolution of her own future risk attitudes. When $\rho =
0$, this channel is absent and the hedging component vanishes entirely, yielding a solution that resembles the one found in Eq. (19) of \cite{BS21}.

The integral term itself involves the elasticity $\frac{\partial_y h^{\bar{y}}(t,y)}{h^{\bar{y}}(t,y)}$, where the function $h^{\bar{y}}$ solves the PIDE in \eqref{eq:PIDE_finalform_main}. As shown in the proof of Proposition \ref{prop:ln-CRRA_preferences} in Appendix \ref{proof_corollary2_application}, this function is tied to the value function of the problem (specifically, to the continuation value), so the hedging demand reflects the sensitivity of future continuation value to the current preference state.

A notable feature of this representation is that, despite the complexity of the underlying dynamic inconsistency, the equilibrium policy does not depend on wealth. This property, of course, follows from the homothetic structure of CRRA preferences and is preserved here even in the presence of evolving risk attitudes. It is worth contrasting this with the findings of \cite{KrausslLucasSiegman2012:FinResearchLetters}, who --working with a different model for preferences-- show that uncertainty about risk aversion can instead generate a positive relation between wealth and risk taking.

\begin{figure}[!t]
\hspace{-3em}
\subfloat[Equilibrium policy, $\mu_Y = 0.02$]{\includegraphics[width=0.6\textwidth]{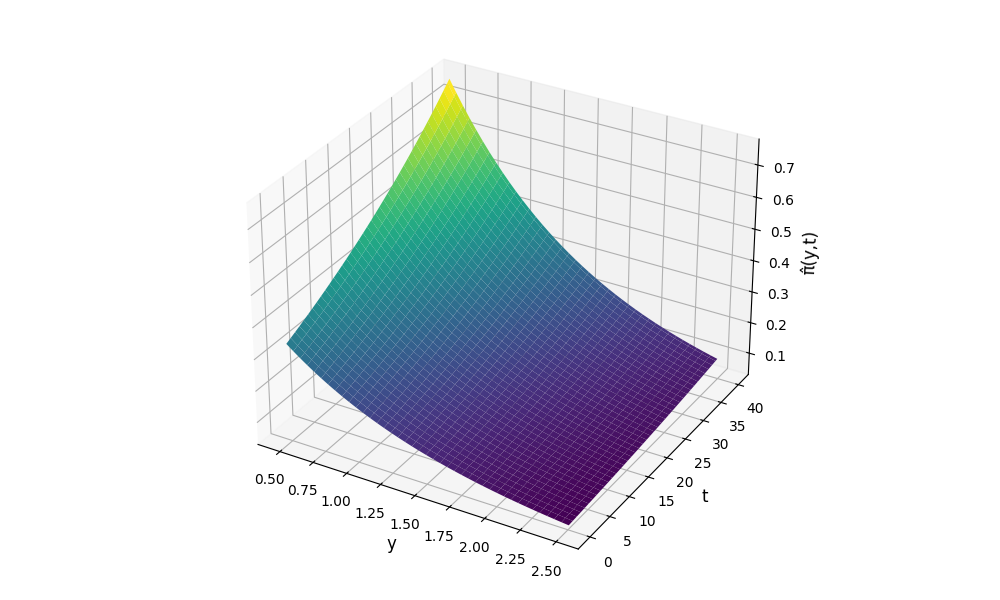}} 
\hspace{-4em}
\subfloat[Equilibrium policy, $\mu_Y = -0.02$]{\includegraphics[width=0.6\textwidth]{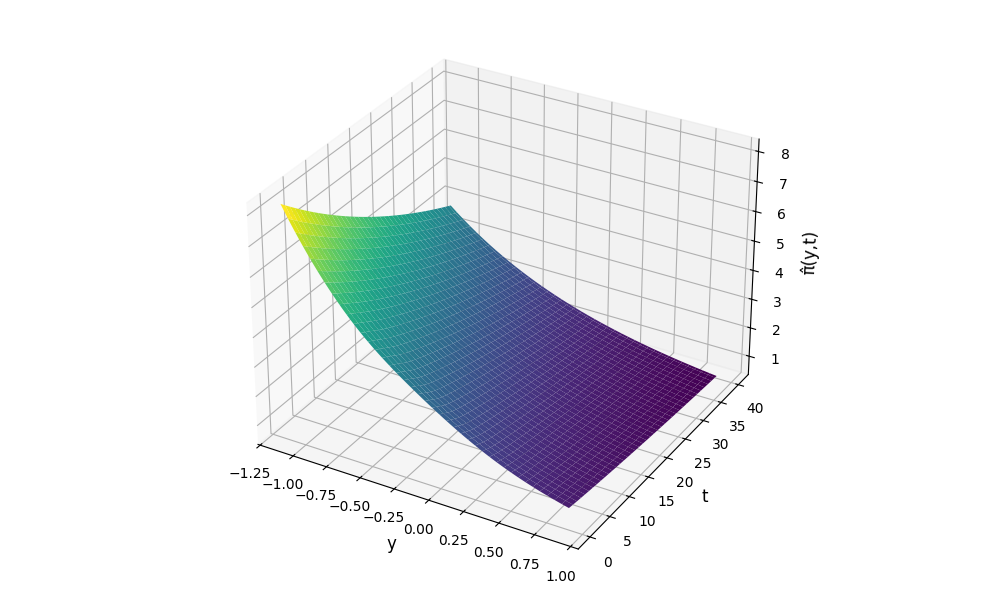}} 
\\

\hspace{-4em}
\subfloat[Preference hedging demand, $\mu_Y = 0.02$]{\includegraphics[width=0.6\textwidth]{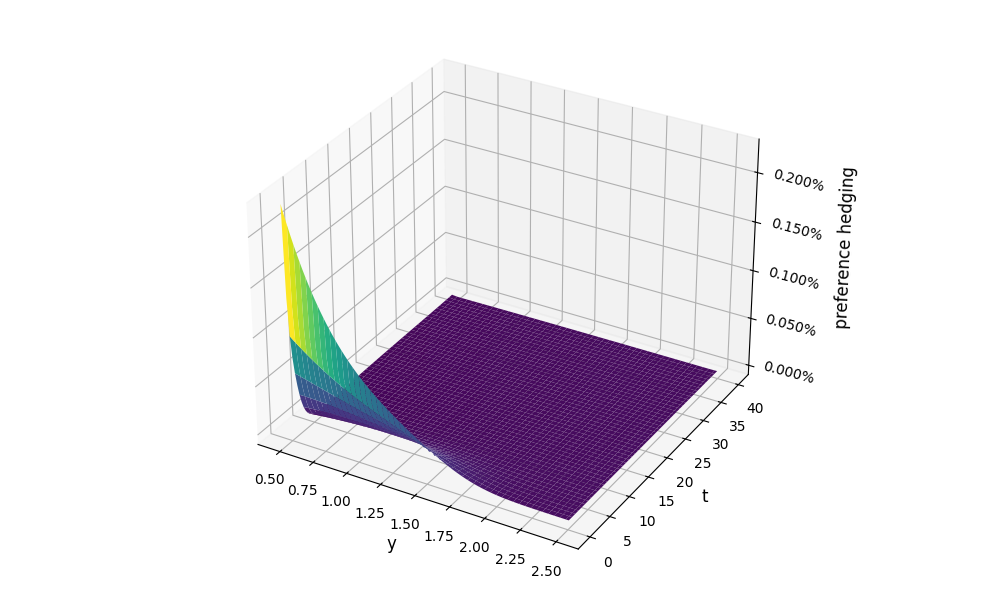}} 
\hspace{-3em}
\subfloat[Preference hedging demand, $\mu_Y = -0.02$]{\includegraphics[width=0.6\textwidth]{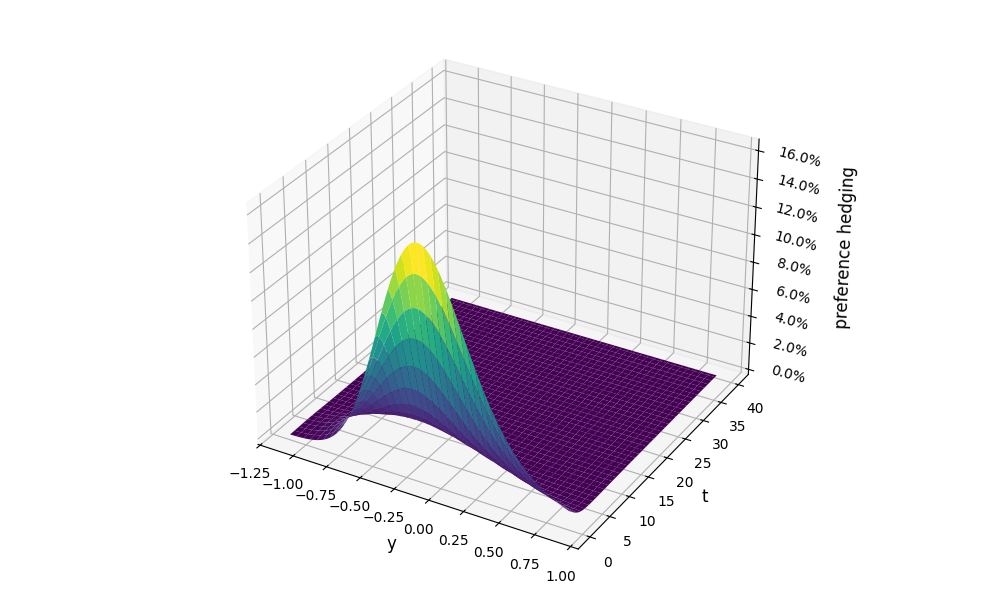}} 
\caption{\footnotesize Equilibrium policy $\widehat{\pi}(t,y)$ and preference hedging demand for  $\mu_Y = 0.02$ (left column) and $\mu_Y = -0.02$ (right column). Other parameters: $(r,\mu_S, \sigma_S, \sigma_{Y}, \rho, \exp(y_0)) = (0.02, \, 0.07, \, 0.2, \, 0.04, \, 0.6, \, 2)$. 
}
\label{fig:EquilibriumStrategy_Plot}
\end{figure}

\medskip

Figures \ref{fig:EquilibriumStrategy_Plot} (a)-(b)  show the equilibrium policy for $\mu_Y = \{-0.02,0.02\}$, while Figures \ref{fig:EquilibriumStrategy_Plot}  (c)-(d) display the corresponding preference hedging demand. Other parameters in the computations are set as follows: $(r,\mu_S, \sigma_S, \sigma_{Y}, \rho, \exp(y_0)) = (0.02, \, 0.07, \, 0.2, \, 0.04, \, 0.6, \, 2)$.

A first observation is the asymmetry between positive and negative $\mu_Y$. When $\mu_Y=0.02$,  the expected drift of future risk aversion is positive. For a fixed current state $y$, the agent behaves more conservatively at early dates, and the equilibrium policy increases over time. The associated preference-hedging demand is positive but small and decays quickly. Its economic role is to partially counteract the conservative initial attitude, although quantitatively the effect is minor.

When $\mu_Y = -0.02$, the expected drift of future risk aversion is negative. The agent initially invests more aggressively, anticipating a less risk-averse future self, and the equilibrium policy decreases over time. Here, the hedging demand reinforces the agent's initial behavior and is larger in magnitude. A plausible explanation is that the distribution of $Y_T$ shifts towards regions where $h^{\bar{y}}$ is more sensitive in $y$, leading to a higher weighted elasticity.

It is important to emphasize that, in standard models of time-inconsistent preferences, agents may foresee changes in utility, but equilibrium strategies typically do not feature an explicit hedging motive against such preference shifts. In our setting, however, a hedging component arises endogenously from the sensitivity structure of the continuation value with respect to the preference factor $Y$.

Additional numerical results, including tables, are provided in Appendix \ref{app:additiona_numerics}. The pseudocode for the computational procedure is given in Appendix \ref{app:pseudocode}.

\section{Conclusion} \label{sec:Conclusions}
We developed a continuous-time investment framework in which risk preferences evolve endogenously with an observable state variable. By evaluating payoffs through conditional certainty equivalents and aggregating them across future preference states, we obtained a reward functional that is inherently time-inconsistent. This required refining existing equilibrium methods for time-inconsistent control problems and deriving a new one suited to our setting.

Specializing to CRRA preferences, we let the underlying preference factor follow an arithmetic Brownian motion and define relative risk aversion as its exponential --yielding \textit{de facto} a geometric Brownian motion for risk aversion itself. Under this specification, we showed that the equilibrium system reduces to a coupled nonlinear PIDE indexed by terminal preference states. The resulting semi-explicit equilibrium portfolio rule features a novel hedging component that captures incentives to adjust current exposure in the risky asset in anticipation of future changes in risk attitudes.

Possible avenues for future work include expanding the market environment (for instance, by introducing incompleteness), incorporating intermediate evaluation of consumption, and allowing for path dependence in the evolution of preferences over time.

\section*{Acknowledgements}
We gratefully thank the institutions at which parts of this work were carried out. Luca De Gennaro Aquino acknowledges the University of Copenhagen, the Johannes Kepler University Linz, and the University of Lausanne for their support and hospitality. Sascha Desmettre is grateful to the University of Copenhagen and to the University of Lausanne for supporting the respective research visits. Yevhen Havrylenko acknowledges the financial support from the PRIME program (\url{https://www.daad.de/en/studying-in-germany/scholarships/daad-funding-programmes/prime/prime-fellows-202324/}) of the German Academic Exchange Service (DAAD, \url{https://ror.org/039djdh30}), funded by the German Federal Ministry of Education and Research (BMBF). He further acknowledges the generous support of the University of Copenhagen and Ulm University, where parts of this research were done, and the Johannes Kepler University Linz for its hospitality during his research visit.

\bibliography{BibFile} 

\appendix

\section*{Appendices}

\section{Proofs}

\subsection{Proof of Lemma \ref{lem:general_state_process_SDEs_under_Pcond}}\label{proof:lemma_different_dynamics}
To derive the dynamics of $Y$  under $\Pop_{t, x, y, \ybaro}$ (i.e., under the condition $Y_T = \ybaro$),  we can directly use the theory of conditional diffusion processes; see \cite{Delyon2006}. 

To derive the dynamics of $X^{\pi}$, again under $\Pop_{t, x, y, \ybaro}$, we first note that, by Girsanov's theorem, we have
\begin{align*}
\left.\frac{d\Pop_{t, x, y, \ybaro}}{d\Pop_{t, x, y}}\right|_{\mathcal{F}_T} = \exp\Biggl(&-\int_t^{T}\nu^1_sdB^1_s -\int_t^{T}\nu^2_sdB^2_s- \frac{1}{2}\int_t^{T}(\nu^1_s)^2ds - \frac{1}{2}\int_t^{T}(\nu^2_s)^2 ds \Biggr),
\end{align*}
where $\nu^1$ and $\nu^2$ are square-integrable processes characterizing the change of measure. Also, the SDEs of standard Brownian motions under $\Pop_{t, x, y, \ybaro}$ are given by
\begin{equation}\label{eq:Girsanov_BM_relation}
d\Bbar^1_s = dB^1_s + \nu^1_sds,\qquad \,d\Bbar^2_s = dB^2_s + \nu^2_sds.
\end{equation}
Since conditioning on $Y_T = \ybaro$ influences only $B^1$, and $B^1 \perp B^2$, we deduce that the dynamics of $B^2$ remains unchanged under $\Pop_{t, x, y, \ybaro}$. Therefore, for a standard Brownian motion $\Bbar^2$ under $\Pop_{t, x, y, \ybaro}$, we have $d\Bbar^2_s = dB^2_s$ and $\nu^2_s = 0$, for all $s \in [t, T]$. To find the correct $\nu_1$, we insert \eqref{eq:Girsanov_BM_relation} in \eqref{eq:Y_SDE_under_Pbar} and get
\begin{align*}
dY_s = \Big(\mu_Y(s, Y_s)  + \sigma_Y^2(s, Y_s) \partial_y \ln \big(p_Y(s, Y_s;{T},\ybaro)\big) + \sigma_Y(s, Y_s)\nu^1_s\Big)ds + \sigma_Y(s, Y_s) dB^1_s.
\end{align*}
As the drift must be equal to $\mu_Y(s, Y_s)$ under $\Pop_{t,x,y}$, and $\sigma_Y(s, Y_s) \neq 0$ almost surely, for all $s \in [t, T]$, we conclude that
\begin{equation*}\label{eq:nu_1}
\nu^1_s = -\sigma_Y(s, Y_s) \partial_y \ln \big(p_Y(s, Y_s;{T},\ybaro)\big). 
\end{equation*}
Inserting \eqref{eq:Girsanov_BM_relation} in the SDE \eqref{eq:X_SDE_under_Pbar} for $X^{\pi}$ under $\Pop_{t,x,y}$ and using the above form of $\nu_1$ specifying the change of measure, we obtain the SDE of $X^{\pi}$ under $\Pop_{t,x,y,\ybaro}$:
\begin{align*}
d X^\pi_s &= X^\pi_s (r+ \pi(s) (\mu_S-r))ds + X^\pi_s\pi(s) \sigma_S \rBrackets{ \rho \rBrackets{d \Bbar^1_s - \nu^1_sds} + \sqrt{1 - \rho^2} d \Bbar^2_s}\\[0.2cm]
&= X^\pi_s \Big(r+ \pi(s) (\mu_S-r) + \pi(s)\sigma_S \rho \, \sigma_Y(s, Y_s) \, \partial_y \ln \big(p_Y(s, Y_s;{T},\ybaro)\big) \Big) ds\\
& \quad + X^\pi_s\pi(s) \sigma_S \rBrackets{ \rho d \Bbar^1_s + \sqrt{1 - \rho^2} d \Bbar^2_s}. 
\end{align*}
\qed

\subsection{Proof of Theorem \ref{th:verification}}
\label{proof_verification_theorem}

\noindent \textit{Proof of (R1)}. Choose an arbitrary but fixed $\ybaro \in \YSet$ and $(t, x, y) \in [0, T) \times \XSet \times \YSet$.  Let $\epi$ be the $\arg \sup$ in \eqref{eq:J_1:EHJB_PDE_1}, which exists by assumption \textit{(C1)}. Applying It\^{o}'s lemma to the function $g^{\ybaro}$ (which belongs to $\mathcal{C}^{1,2, 2}\left(\TSet \times \XSet \times \YSet \right)$ by assumption \textit{(C3)}) and the state process $(X^{\epi}, Y)$ under the measure $\Pop_{t,x,y, \ybaro}$, we get
\begin{equation}\label{eq:R1_proof_gyy_inttT}
\begin{aligned}
\gybar{T, X^{\epi}_T, Y_T} &= \gybar{t, X^{\epi}_t,  Y_t} + \inttT \Dbarepi \gybar{s, X^{\epi}_s,  Y_s}\,ds \\ 
& \quad + \inttT \partial_{x} \gybar{s, X^{\epi}_s,  Y_s} \epi(s)\sigma_S \rBrackets{ \rho d \Bbar^1_s + \sqrt{1 - \rho^2} d \Bbar^2_s} \\
& \quad + \inttT \partial_{y} \gybar{s, X^{\epi}_s,  Y_s} \sigma_Y d\Bbar^1_s.
\end{aligned}
\end{equation}
Taking the expectation on both sides of \eqref{eq:R1_proof_gyy_inttT}, and using the fact that $\gybar{t,x, y}$ satisfies \eqref{eq:J_inf:EHJB_PDE_2} by assumption \textit{(C2)}, as well as $\gybar{t,x, y} \in \mathcal{L}^{2}\rBrackets{X^{\epi}, Y}$ by assumption \textit{(C4)}, gives
\begin{equation*}
\Eop_{t,x,y,\ybaro}\sBrackets{\gybar{T, X^{\epi}_T, Y_T}} = \Eop_{t,x,y,\ybaro}\sBrackets{\gybar{t, X^{\epi}_t, Y_t}}.
\end{equation*}
Thus, since $\gybar{T,x, y}$ solves \eqref{eq:J_inf:terminal_condition_g} (again by \textit{(C2)}), we have
\begin{equation*}
\gybar{t, x, y} = \Eop_{t,x,y, \ybaro}\sBrackets{\gybar{T, X^{\epi}_T, Y_T}} = \Eop_{t,x,y, \ybaro}\sBrackets{u^{\gamma(Y_T)}\rBrackets{X^{\epi}_T}},
\end{equation*}
which proves \textit{(R1)}.

\bigskip

\noindent \textit{Proof of (R2)}.  Plugging $\epi$ in \eqref{eq:J_inf:EHJB_PDE_1}, we obtain
\begin{align*}
0 &= \mathcal{D}^{\epi} V(t, x, y) - \mathcal{D}^{\epi} \int_\mathcal{Y} \varphi^{\ybaro}\rBrackets{\gybar{t, x, y}} dF_{Y_T}(\ybaro; t, y) \\
& \qquad \qquad \qquad  + \int_\mathcal{Y} \rBrackets{\varphi^{\ybaro}}'\rBrackets{\gybar{t, x, y}} \, \Dbarepi \gybar{t, x, y} \,  dF_{Y_T}(\ybaro; t, y).
\end{align*}
Since \eqref{eq:J_inf:EHJB_PDE_2} holds, the last term in the above PDE is zero. Thus,
\begin{equation}\label{eq:equal_terms_in_HJB_PDE}
\Depi V(t, x, y) = \Depi \int_\mathcal{Y} \varphi^{\ybaro}\rBrackets{\gybar{t, x, y}}  dF_{Y_T}(\ybaro; t, y).
\end{equation}
Due to \textit{(C3)}, $V \in \mathcal{C}^{1,2, 2}\left(\TSet \times \XSet \times \YSet\right)$. Applying It\^{o}'s lemma to $V(t, X^{\epi}_t,  Y_t)$ on $[t, T]$ under the measure $\mathbb{P}_{t, x, y}$, and then taking the expectation on both sides of the equality, we derive that
\begin{align*}
\mathbb{E}_{t, x, y}&\sBrackets{V(T, X^{\epi}_T,  Y_T)} =  V(t, x, y) + \mathbb{E}_{t, x, y}\sBrackets{\inttT \mathcal{D}^{\epi}V(s, X^{\epi}_s,  Y_s)\,ds} \\
& + \mathbb{E}_{t, x, y}\sBrackets{\inttT \partial_{x} V(s, X^{\epi}_s,  Y_s) \epi(s)\sigma_S \rBrackets{ \rho dB^1_s + \sqrt{1 - \rho^2} d B^2_s} } \\
& + \mathbb{E}_{t, x, y}\sBrackets{\inttT \partial_{y} V(s, X^{\epi}_s,  Y_s) \sigma_Y d B^1_s}\\
& \stackrel{(C4)}{=} V(t, x, y) + \mathbb{E}_{t, x, y}\sBrackets{\inttT \mathcal{D}^{\epi}V(s, X^{\epi}_s,  Y_s)\,ds} \\
& \stackrel{\eqref{eq:equal_terms_in_HJB_PDE}}{=}   V(t, x, y) + \mathbb{E}_{t, x, y}\sBrackets{\inttT \Depi \rBrackets{ \int_\mathcal{Y} \varphi^{\ybaro}\rBrackets{\gybar{s, X^{\epi}_s,  Y_s}}  dF_{Y_T}\rBrackets{\ybaro; s, Y_s}}ds}.
\end{align*}
Using the notation for $G_{\infty}(t,x,y)$ in \eqref{eq:G_infty},
we consider the process
\begin{equation*}
G_{\infty}(s, X^{\epi}_s,  Y_s) = \int_\mathcal{Y} \varphi^{\ybaro}\rBrackets{\gybar{s, X^{\epi}_s,  Y_s}} dF_{Y_T}\rBrackets{\ybaro; s, Y_s}.
\end{equation*}
Due to \textit{(C3)}, $G_{\infty} \in \mathcal{C}^{1,2, 2}\left(\TSet \times \XSet \times \YSet\right)$. Applying It\^{o}'s lemma to $G_{\infty}$ under $\Pop_{t,x,y}$, taking the expectation and using that $G_{\infty} \in \mathcal{L}^{2}\rBrackets{X^{\epi}, Y}$ (again by \textit{(C4)}), we get
\begin{align*}
\mathbb{E}_{t, x, y}&\sBrackets{G_{\infty}(T, X^{\epi}_T,  Y_T)} -  \mathbb{E}_{t, x, y}\sBrackets{G_{\infty}(t, X^{\epi}_t,  Y_t)}\\
& = \mathbb{E}_{t, x, y}\sBrackets{\inttT \Depi \rBrackets{ \int_\mathcal{Y} \varphi^{\ybaro}\rBrackets{\gybar{s, X^{\epi}_s,  Y_s}} dF_{Y_T}\rBrackets{\ybaro; s, Y_s}}ds}.
\end{align*}
Therefore,
\begin{align*}
\mathbb{E}_{t, x, y}\sBrackets{V(T, X^{\epi}_T,  Y_T)} &= V(t, x, y) + \mathbb{E}_{t, x, y}\sBrackets{G_{\infty}(T, X^{\epi}_T,  Y_T)} - G_{\infty}(t, X^{\epi}_t,  Y_t).
\end{align*}
Due to \eqref{eq:J_inf:terminal_condition_V} and \eqref{eq:J_inf:terminal_condition_g}, 
$\mathbb{E}_{t, x, y}\sBrackets{V(T, X^{\epi}_T,  Y_T)} = \mathbb{E}_{t, x, y}\sBrackets{G_{\infty}(T, X^{\epi}_T,  Y_T)}$. Thus,
\begin{align*}
V(t, x, y) & = G_{\infty}(t, x,  y) = \int_\mathcal{Y} \varphi^{\ybaro}\rBrackets{\gybar{t, x,  y}} dF_{Y_T}\rBrackets{\ybaro; t, y}\\
& \stackrel{(R1)}{=}  \int_\mathcal{Y} \varphi^{\ybaro}\rBrackets{\Eop_{t,x,y, \ybaro}\sBrackets{u^{\gamma(Y_T)}\rBrackets{X^{\epi}_T}}} dF_{Y_T}\rBrackets{\ybaro; t, y}  \\
& = \Eop_{t,x, y}\left[ v \circ \left(u^{\gamma(Y_T)}\right)^{-1}\left(\Eop_{t,x,y, \ybaro}\left[u^{\gamma(Y_T)}(X^{\pi}_T)\right]\right) \right] \\
& \stackrel{\eqref{eq:reward_functional}}{=} J^\pi(t,x,y),
\end{align*}
which proves \textit{(R2).}

\bigskip

\noindent \textit{Proof of (R3).} First, we derive a recursive representation of $\gybar{t, x,  y}$ and $J^{\pi_\delta}(t, x, y)$ for an arbitrary but fixed $\pi_\delta$. Similarly to \eqref{eq:R1_proof_gyy_inttT}, under the measure $\Pop_{t, x, y, \ybaro}$, we can apply It\^{o}'s lemma to the process $\gybar{s, X^{\pi_\delta}_s,  Y_s}$ on the time interval $[t+\delta, T]$:
\begin{equation}\label{eq:R3_proof_gyy_intthT}
\begin{aligned}
\gybar{T, X^{\pi_\delta}_T, Y_T} &= \gybar{t + \delta, X^{\pi_\delta}(t+\delta),  Y(t+\delta)} + \intthT \overline{D}^{\pi_\delta} \gybar{s, X^{\pi_\delta}_s,  Y_s}\,ds \\ 
& \quad + \intthT \partial_{x} \gybar{s, X^{\pi_\delta}_s,  Y_s} \pi_\delta(s)\sigma_S \rBrackets{ \rho d \Bbar^1_s + \sqrt{1 - \rho^2} d \Bbar^2_s} \\
& \quad + \intthT \partial_{y} \gybar{s, X^{\pi_\delta}_s,  Y_s} \sigma_Y d\Bbar^1_s.
\end{aligned}
\end{equation}
Taking the expectation and using $\pi_\delta(s) = \epi(s)$ , for every  $s \in [t + \delta, T]$, together with assumptions \textit{(C2)} and \textit{(C4)}, we get
\begin{equation} \label{eq:R3_proof_gybar_th}
\Eop_{t, x, y, \ybaro}\sBrackets{ \gybar{t + \delta, X^{\pi_\delta}(t+\delta),  Y(t+\delta)} } = \Eop_{t,x,y, \ybaro}\sBrackets{u^{\gamma(\ybaro)}\rBrackets{X^{\pi_\delta}_T}}.
\end{equation}
Furthermore, we have
\begin{align}
J^{\pi_\delta}(t,x,y) & \stackrel{\eqref{eq:reward_functional_explicit}}{=}  \int_\mathcal{Y} \varphi^{\ybaro} \rBrackets{\Eop_{t, x, y, \ybaro} \sBrackets{u^{\gamma(\ybaro)}\rBrackets{X^{\pi_\delta}_T}}} \, dF_{Y_T}(\ybaro; t, y) \notag \\
& = \int_\mathcal{Y} \varphi^{\ybaro} \rBrackets{\Eop_{t, x, y, \ybaro} \sBrackets{u^{\gamma(\ybaro)}\rBrackets{X^{\pi_\delta}_T}}} \, dF_{Y_T}(\ybaro; t, y) \label{eq:R3_proof_J_pih_recursion_before_E} \\
& \quad + V(t + \delta, X^{\pi_\delta}(t+\delta), Y(t+\delta)) - J^{\pi_\delta}(t + \delta, X^{\pi_\delta}(t+\delta), Y(t+\delta)),\notag
\end{align}
where the difference between the last two terms is zero due to $\pi_\delta(s) = \epi(s), s \in [t+\delta,T]$:
\begin{equation*}
\begin{aligned}
J^{\pi_\delta}(t + \delta, X^{\pi_\delta}(t+\delta), Y(t+\delta)) & \; = J^{\epi}(t + \delta, X^{\pi_\delta}(t+\delta), Y(t+\delta)) \\
& \stackrel{\textit{(R2)}}{=} V(t + \delta, X^{\pi_\delta}(t+\delta), Y(t+\delta)).
\end{aligned}
\end{equation*}
Taking the expectation under the measure $\Pop_{t, x, y}$ in \eqref{eq:R3_proof_J_pih_recursion_before_E} yields
\begin{equation} \label{eq:R3_proof_J_pih_recursion_after_E}
\begin{aligned}
J^{\pi_\delta}(t,x,y) & = \int_\mathcal{Y} \varphi^{\ybaro} \rBrackets{\Eop_{t, x, y, \ybaro} \sBrackets{u^{\gamma(\ybaro)}\rBrackets{X^{\pi_\delta}_T}}} \, dF_{Y_T}(\ybaro; t, y)  \\
& \quad + \Eop_{t, x, y}\sBrackets{V(t + \delta, X^{\pi_\delta}(t+\delta), Y(t+\delta))} \\
& \quad - \Eop_{t, x, y}\sBrackets{J^{\pi_\delta}(t + \delta, X^{\pi_\delta}(t+\delta), Y(t+\delta))}.
\end{aligned}
\end{equation}
Therefore, 
\begin{align}
J^{\epi}(t,x,y)  - J^{\pi_\delta}(t,x,y)  & \stackrel[]{\eqref{eq:R3_proof_J_pih_recursion_after_E}+\textit{(R2)}}{=}V(t, x, y) - \Eop_{t, x, y}\sBrackets{V(t + \delta, X^{\pi_\delta}(t+\delta), Y(t+\delta))} \notag \\
& \quad - \int_\mathcal{Y} \varphi^{\ybaro} \rBrackets{\Eop_{t, x, y, \ybaro} \sBrackets{u^{\gamma(\ybaro)}\rBrackets{X^{\pi_\delta}_T}}} \, dF_{Y_T}(\ybaro; t, y)  \notag \\
& \quad + \Eop_{t, x, y}\sBrackets{J^{\pi_\delta}(t + \delta, X^{\pi_\delta}(t+\delta), Y(t+\delta))} \notag \\[0.2cm]
& = - \rBrackets{\Eop_{t, x, y}\sBrackets{V(t + \delta, X^{\pi_\delta}(t+\delta), Y(t+\delta))} - V(t, x, y)} \notag \\
& \quad - \Bigl(\int_\mathcal{Y} \varphi^{\ybaro} \rBrackets{\Eop_{t, x, y, \ybaro}\sBrackets{ \gybar{t + \delta, X^{\pi_\delta}(t+\delta),  Y(t+\delta)} }} \, dF_{Y_T}(\ybaro; t, y) \notag \\
& \quad \quad  - \int_\mathcal{Y} \varphi^{\ybaro} \rBrackets{\gybar{t, x, y}} \, dF_{Y_T}(\ybaro; t, y) \Bigr) \notag \\
& \quad + \Bigl(\Eop_{t, x, y}\sBrackets{J^{\pi_\delta}(t + \delta, X^{\pi_\delta}(t+\delta), Y(t+\delta))} \notag\\
& \quad \quad -\int_\mathcal{Y} \varphi^{\ybaro} \rBrackets{\gybar{t, x, y}} \, dF_{Y_T}(\ybaro; t, y) \Bigr) \notag \\
& =: -\Delta_1(\delta)  - \Delta_2(\delta) + \Delta_3(\delta), \label{eq:R3_proof_def_delta_terms} 
\end{align}
where in the second equality we add and subtract the same term and use \eqref{eq:R3_proof_gybar_th} for $\Eop_{t,x,y, \ybaro}\sBrackets{u^{\gamma(\ybaro)}\rBrackets{X^{\pi_\delta}_T}}$.

To prove that $\epi$ is an equilibrium control according to Definition \ref{def:equi_control}, we need to compute the following limit:
\begin{align}
\liminf_{\delta \downarrow 0} &\frac{J^{\hat\pi}(t,x,y) - J^{\pi_\delta}(t,x,y)}{\delta} \stackrel{\eqref{eq:R3_proof_def_delta_terms}}{=} \liminf_{\delta \downarrow 0}\frac{1}{\delta}\rBrackets{-\Delta_1(\delta) - \Delta_2(\delta) +\Delta_3(\delta)} \notag \\
& = - \rBrackets{\limsup_{\delta \downarrow 0}\frac{1}{\delta}\Delta_1(\delta) + \limsup_{\delta \downarrow 0}\frac{1}{\delta}\Delta_2(\delta) - \liminf_{\delta \downarrow 0}\frac{1}{\delta}\Delta_3(\delta)},\label{eq:epi_liminf_via_limsup}
\end{align}
where we use the linearity of the limit and the relation between $\limsup$ and $\liminf$.

Using standard arguments, we can show that
\begin{equation}\label{eq:R3_proof_limit_1} 
\limsup_{\delta \downarrow 0} \frac{1}{\delta}\Delta_1(\delta) = \mathcal{D}^{\pi}V(t, x, y).
\end{equation}
Furthermore, we get
\begin{align}
\limsup_{\delta \downarrow 0}& \frac{1}{\delta} \Delta_2(\delta) =  \limsup_{\delta \downarrow 0} \frac{1}{\delta} \Biggl( \int_\mathcal{Y} \varphi^{\ybaro} \rBrackets{\Eop_{t, x, y, \ybaro}\sBrackets{ \gybar{t + \delta, X^{\pi_\delta}(t+\delta),  Y(t+\delta)} }} \, dF_{Y_T}(\ybaro; t, y) \notag \\
& \quad \quad  - \int_\mathcal{Y} \varphi^{\ybaro} \rBrackets{\gybar{t, x, y}} \, dF_{Y_T}(\ybaro; t, y) \Biggr) \notag \\
& = \int_\mathcal{Y} \limsup_{\delta \downarrow 0} \frac{1}{\delta}\Biggl(\varphi^{\ybaro} \rBrackets{\Eop_{t, x, y, \ybaro}\sBrackets{ \gybar{t + \delta, X^{\pi_\delta}(t+\delta),  Y(t+\delta)} }}  \notag \\
& \quad \quad  - \varphi^{\ybaro} \rBrackets{\gybar{t, x, y}} \Biggr)\, dF_{Y_T}(\ybaro; t, y) \notag \\
& =  \int_\mathcal{Y} \limsup_{\delta \downarrow 0} \Biggl(\frac{\varphi^{\ybaro} \rBrackets{\Eop_{t, x, y, \ybaro}\sBrackets{ \gybar{t + \delta, X^{\pi_\delta}(t+\delta),  Y(t+\delta)} }} - \varphi^{\ybaro} \rBrackets{\gybar{t, x, y}}}{\Eop_{t, x, y, \ybaro}\sBrackets{ \gybar{t + \delta, X^{\pi_\delta}(t+\delta),  Y(t+\delta)} } - \gybar{t, x, y}}  \notag \\
& \quad \quad \cdot  \frac{\Eop_{t, x, y, \ybaro}\sBrackets{ \gybar{t + \delta, X^{\pi_\delta}(t+\delta),  Y(t+\delta)} } - \gybar{t, x, y}}{\delta} \Biggr)\, dF_{Y_T}(\ybaro; t, y) \notag \\ 
& =  \int_\mathcal{Y} \rBrackets{\varphi^{\ybaro}}'\rBrackets{ \gybar{t, x, y}}  \,  \overline{\mathcal{D}}^{\pi} \gybar{t, x, y} \, dF_{Y_T}(\ybaro; t, y) \notag \\
& \hspace{-0.2cm} \stackrel{\eqref{eq:barH_pi}}{=} \overline{\mathcal{H}}^{\pi}\rBrackets{\rBrackets{g^{\ybaro}}_{\ybaro \in \YSet}}(t, x, y),
\label{eq:R3_proof_limit_2}
\end{align}
where in the second equality we use regularity conditions \textit{(C4)} to exchange $\liminf$ and integral, in the third equality we multiply and divide by the same non-zero term, and in the fourth equality we use the definition of a derivative for the first term, the definition of the differential operator under $\Pop_{t,x,y,\ybaro}$, the assumption that $\varphi^{\ybaro}$ is differentiable, the condition that $g^{\ybaro} \in \overline{\mathcal{L}}^2(X^{\pi}, Y)$ as per \textit{(C4)}, and the product rule for limits.

Finally, observing that
\begin{align*}
J^{\pi_\delta}&(t + \delta, X^{\pi_\delta}(t+\delta), Y(t+\delta)) \\
& \stackrel{\eqref{eq:reward_functional_explicit}}{=} \int_\mathcal{Y} \varphi^{\ybaro} \rBrackets{\Eop_{t + \delta, X^{\pi_\delta}(t+\delta), Y(t+\delta), \ybaro} \sBrackets{u^{\gamma(\ybaro)}\rBrackets{X^{\pi}_T}}} \, dF_{Y_T}(\ybaro; t + \delta, Y(t+\delta)) \\
& \stackrel{(R1)}{=} \int_\mathcal{Y} \varphi^{\ybaro} \rBrackets{\gybar{t + \delta, X^{\pi_\delta}(t+\delta), Y(t+\delta)}} \, dF_{Y_T}(\ybaro; t + \delta, Y(t+\delta)) \\
& \stackrel{\eqref{eq:G_infty}}{=}  G_{\infty}(t + \delta, X^{\pi_\delta}(t+\delta), Y(t+\delta)),
\end{align*}
we compute
\begin{align}
\liminf_{\delta \downarrow 0} \frac{1}{\delta}\Delta_3(\delta) &= \liminf_{\delta \downarrow 0} \frac{1}{\delta}  \Biggl(\Eop_{t, x, y}\sBrackets{J^{\pi_\delta}(t + \delta, X^{\pi_\delta}(t+\delta), Y(t+\delta))} \notag \\
& \quad \quad -\int_\mathcal{Y} \varphi^{\ybaro} \rBrackets{\gybar{t, x, y}} \, dF_{Y_T}(\ybaro; t, y) \Biggr) \notag \\
& \stackrel{\eqref{eq:G_infty}}{=} \liminf_{\delta \downarrow 0} \frac{1}{\delta}  \Biggl(\Eop_{t, x, y}\sBrackets{ G_{\infty}(t + \delta, X^{\pi_\delta}(t+\delta), Y(t+\delta)) } - G_{\infty}(t, x, y)\Biggr) \notag \\
& \stackrel{}{=} \mathcal{D}^{\pi}G_{\infty}(t, x, y).\label{eq:R3_proof_limit_3}
\end{align}
Using \eqref{eq:R3_proof_def_delta_terms} and the convergence results in \eqref{eq:R3_proof_limit_1}-\eqref{eq:R3_proof_limit_3}, we obtain
\begin{align*}
\liminf_{\delta \downarrow 0} \, &\frac{J^{\hat\pi}(t,x,y) - J^{\pi_\delta}(t,x,y)}{\delta} \\
& = - \rBrackets{\limsup_{\delta \downarrow 0}\frac{1}{\delta}\Delta_1(\delta) + \limsup_{\delta \downarrow 0}\frac{1}{\delta}\Delta_2(\delta) - \liminf_{\delta \downarrow 0}\frac{1}{\delta}\Delta_3(\delta)} \\
& = - \rBrackets{ \mathcal{D}^{\pi}V(t, x, y)  + \overline{\mathcal{H}}^{\pi}\rBrackets{\rBrackets{g^{\ybaro}}_{\ybaro \in \YSet}}(t, x, y) - \mathcal{D}^{\pi}G_{\infty}(t, x, y)} \\
& \geq 0,
\end{align*}
where the inequality follows from the fact that $\epi$ realizes the supremum in \eqref{eq:J_inf:EHJB_PDE_1} as per \textit{(C1)}. Therefore, $\epi$ is an equilibrium control, which proves \textit{(R3)}.

\bigskip

\noindent \textit{Proof of (R4).} 
We conclude that $V(t,x,y)$ is indeed the equilibrium value function, i.e., $\widehat{V}(t, x, y) =  V(t,x,y)$, since $V(t,x,y) = J^{\epi}(t,x,y)$ by \textit{(R2)} and $\epi$ is an equilibrium control by \textit{(R3)}. This proves \textit{(R4)} and completes the proof of the verification theorem. \qed

\subsection{Proof of Corollary \ref{cor:special_case:state_process_SDEs_under_Pcond}}
\label{proof_corollary1_application}

First, it is straightforward to verify that 
\begin{equation*}
    f_{Y_T}(\ybaro; t, y) = \frac{1}{\sqrt{2 \uppi  \sigma_Y^2 (T - t)}} \exp\rBrackets{-\frac{\rBrackets{\ybaro - y - \mu_Y (T - t)}^2}{2 \sigma_Y^2 (T - t)}}.
\end{equation*}
(Note that $\uppi$ in the conditional PDF above denotes the mathematical constant pi, so it should not be confused with the control (investment strategy) denoted by $\pi$.)

Thus, for $Y_T|Y_s = y$, we obtain
\begin{equation}
\label{eq:dy_ln_p_special_case}
\begin{split}
\partial_y \ln \big(p_Y(s, y;T,\ybaro)\big) &= \partial_y \ln \rBrackets{\frac{1}{\sqrt{2 \uppi \sigma_Y^2 (T-s)}}\exp\rBrackets{-\frac{\rBrackets{\ybaro - (y + \mu_Y(T-s))}^2}{2 \sigma_Y^2 (T - s)}}} \\
& = -\frac{\ybaro - (y + \mu_Y(T-s))}{\sigma_Y^2 (T - s)}. 
\end{split}
\end{equation}
Plugging this result in \eqref{eq:Y_SDE_under_Pbar}, we obtain the SDE for $\rBrackets{Y_s}_{s \in [t, T)}$ under the condition $Y_T = \ybaro$:
\begin{align*}
dY_s & = \rBrackets{\mu_Y ds  + \sigma_Y^2 \rBrackets{-\frac{\ybaro - (y + \mu_Y(T-s))}{\sigma_Y^2 (T - s)}}}ds + \sigma_Y dB^1_s\\
& = \frac{\ybaro - Y_s}{T - s}ds + \sigma_Y d \Bbar^1_s,
\end{align*}
with $Y_t = y$.

\medskip

Similarly, inserting \eqref{eq:dy_ln_p_special_case} into \eqref{eq:X_SDE_under_Pbar}, using that $\sigma_Y(s,Y_S) = \sigma_Y$, and simplifying, we obtain the SDE for $(X^{\pi})_{s \in [t,T)}$:
\begin{equation*}
\begin{aligned}
dX^\pi_s & = X^\pi_s \rBrackets{r+ \pi(s) (\mu_S-r) + \pi(s)\rho \frac{\sigma_S }{\sigma_Y} \frac{\ybaro - Y_s - \mu_Y(T - s)}{T - s}}ds\\
& \qquad + X^\pi_s\pi(s) \sigma_S \rBrackets{ \rho d \Bbar^1_s + \sqrt{1 - \rho^2} d \Bbar^2_s},
\end{aligned}
\end{equation*}
with $X^\pi_t = x$ and $X^\pi_T = \lim_{t \to T} X^\pi_t$.
\qed

\subsection{Proof of Proposition \ref{prop:ln-CRRA_preferences}}

\label{proof_corollary2_application}

Recall our choice of the CRRA specification $u^{\ybaro}(x) = \frac{1}{1 - \gamma(\ybaro)} x^{1 - \gamma(\ybaro)}$ and the aggregator $v(x) = \ln(x)$. The inverse utility and certainty equivalent transform then satisfy
\begin{equation*}
\begin{aligned}
    \rBrackets{u^{\ybaro}}^{-1}(x) &= \big(\rBrackets{1 - \gamma(\ybaro)} x\big)^{\frac{1}{1 - \gamma(\ybaro)}}, \\
    \varphi^{\ybaro}(x) & = \ln\rBrackets{\big(\rBrackets{1 - \gamma(\ybaro)} x\big)^{\frac{1}{1 - \gamma(\ybaro)}}} = \frac{1}{1 - \gamma(\ybaro)}\big(\ln\rBrackets{1 - \gamma(\ybaro)} + \ln(x)\big),\\
    \rBrackets{\varphi^{\ybaro}}'(x) &= \frac{1}{x\rBrackets{1 - \gamma(\ybaro)}}.
\end{aligned}
\end{equation*}
Using these expressions and the evolution of $(X^{\pi}, Y)$, the  in \eqref{EHJB_eq_1}-\eqref{EHJB_eq_3} takes the form
\begin{equation*}
\begin{split}
&0 =  \sup_{\pi} \int_{\mathcal{Y}} \dfrac{1}{g^{\ybaro}(t, x, y)(1-\gamma(\bar{y}))} \Bigg( \partial_t g^{\ybaro}(t, x, y) \Bigg. \\
& \hspace{2cm}  + x\left(r+\pi\left(\mu_S-r + \rho \dfrac{\sigma_S}{\sigma_{Y}} \dfrac{\bar{y} - y - \mu_Y(T-t) }{T-t} \right) \right) \partial_x g^{\ybaro}(t, x, y)  \\
& \hspace{2cm} + \dfrac{\bar{y}-y}{T-t} \partial_y g^{\ybaro}(t, x, y) + \dfrac{1}{2}x^2 \pi^2 \sigma_S^2 \partial_{xx} g^{\ybaro}(t, x, y)  + \dfrac{1}{2}\sigma_{Y}^2 \partial_{yy} g^{\ybaro}(t, x, y) \\
& \hspace{2cm}\Bigg. + \rho x \pi \sigma_S \sigma_{Y} \partial_{xy} g^{\ybaro}(t, x, y) \Bigg) f_{Y_T}(\bar{y}; t,y) d\bar{y}, \\
& 0  = \partial_t g^{\ybaro}(t, x, y) + x\left(r+\widehat{\pi}\left(\mu_S-r + \rho \dfrac{\sigma_S}{\sigma_{Y}} \dfrac{\bar{y} - y - \mu_Y(T-t) }{T-t} \right) \right) \partial_x g^{\ybaro}(t, x, y)  \\
 & \qquad + \dfrac{\bar{y}-y}{T-t} \partial_y g^{\ybaro}(t, x, y) + \dfrac{1}{2}x^2 \widehat{\pi}^2 \sigma_S^2 \partial_{xx} g^{\ybaro}(t, x, y)  + \dfrac{1}{2}\sigma_{Y}^2 \partial_{yy} g^{\ybaro}(t, x, y) \\
 & \qquad + \rho x \widehat{\pi} \sigma_S \sigma_{Y} \partial_{xy} g^{\ybaro}(t, x, y), \\
  u^{\gamma(\bar{y})}(x) & = g^{\ybaro}(T, x, y). 
\end{split}
\end{equation*}
The first order condition for $\pi$ gives
\begin{equation*}
\begin{split}
    & \widehat{\pi}(t,x,y) = \dfrac{1}{\sigma_S^2 x^2}\int_\mathcal{Y} \dfrac{1}{g^{\ybaro}(t, x, y)\big(1-\gamma(\bar{y})\big)} \bigg(\left(-\mu_S+r - \rho \dfrac{\sigma_S}{\sigma_{Y}} \dfrac{\bar{y} - y - \mu_Y(T-t) }{T-t} \right) \partial_x g^{\ybaro}(t, x, y) \bigg. \Bigg. \\
   & \Bigg. \bigg. \hspace{4cm} - \rho \sigma_S \sigma_{Y} \partial_{xy} g^{\ybaro}(t, x, y)  \bigg) f_{Y_T}(\bar{y}; t,y) d \bar{y}   \\
   & \hspace{2cm}\times \Bigg( \int_\mathcal{Y} \dfrac{1}{g^{\ybaro}(t, x, y)\big(1-\gamma(\bar{y})\big)} \partial_{xx} g^{\ybaro}(t, x, y) f_{Y_T}(\bar{y}; t,y) d \bar{y}\Bigg)^{-1}.
  \end{split}  
\end{equation*}
We now apply the ansatz 
\begin{equation*}
    g^{\bar{y}}(t,x,y) = \dfrac{1}{1-\gamma(\bar{y})}h^{\bar{y}}(t,y) x^{1-\gamma(\bar{y})},
\end{equation*}
which leads to the following PIDE:
\begin{equation*}
    \begin{split}
        0 & = \dfrac{1}{1-\gamma(\bar{y})} x^{1-\gamma(\bar{y})} \partial_t h^{\bar{y}}(t,y) \\
        & + x\left(r+\widehat{\pi}\left(\mu_S-r + \rho \dfrac{\sigma_S}{\sigma_{Y}} \dfrac{\bar{y} - y - \mu_Y(T-t) }{T-t} \right) \right) x^{-\gamma(\bar{y})} h^{\bar{y}}(t,y)  \\
 & \qquad + \dfrac{\bar{y}-y}{T-t} \left(\dfrac{1}{1-\gamma(\bar{y})} x^{1-\gamma(\bar{y})} \right) \partial_y h^{\bar{y}}(t,y) - \dfrac{1}{2}x^2 \widehat{\pi}^2 \sigma_S^2 \, \gamma(\bar{y})  x^{-\gamma(\bar{y})-1} h^{\bar{y}}(t,y) \\
 & \qquad + \dfrac{1}{2}\sigma_{Y}^2 \left(\dfrac{1}{1-\gamma(\bar{y})} x^{1-\gamma(\bar{y})} \right) \partial_{yy} h^{\bar{y}}(t,y) + \rho x \widehat{\pi} \sigma_S \sigma_{Y} x^{-\gamma(\bar{y})} \partial_{y} h^{\bar{y}}(t,y),
    \end{split}
\end{equation*}
with terminal condition $h^{\bar{y}}(T,y) = 1.$ 

Simplifying with respect to $x$ (which cancels out entirely), this reduces to
\begin{equation*}
\begin{split}
& 0 = \partial_t h^{\bar{y}}(t,y) + \left(r+ \widehat{\pi}\left(\mu_S-r + \rho \dfrac{\sigma_S}{\sigma_{Y}} \dfrac{\bar{y}-y-\mu_Y(T-t)}{T-t} \right) \right) \big(1-\gamma(\bar{y})\big) h^{\bar{y}}(t,y) \\
& \qquad + \dfrac{\bar{y}-y}{T-t} \partial_y h^{\bar{y}}(t,y) - \dfrac{1}{2} \widehat{\pi}^2 \sigma_S^2  \, \gamma(\bar{y}) \big(1-\gamma(\bar{y})\big) h^{\bar{y}}(t,y) + \dfrac{1}{2}\sigma_{Y}^2 \partial_{yy} h^{\bar{y}}(t,y)  \\
& \qquad + \rho \widehat{\pi} \sigma_S \sigma_{Y}  \big(1-\gamma(\bar{y})\big) \partial_y h^{\bar{y}}(t,y).
\end{split}
\end{equation*}
Collecting terms near partial derivatives, we then have
\begin{equation*} 
    \begin{split}
& 0 = \partial_t h^{\bar{y}}(t,y) + \left( \dfrac{\bar{y}-y}{T-t} + \rho \widehat{\pi} \sigma_S \sigma_{Y}  \big(1-\gamma(\bar{y})\big) \right) \partial_y h^{\bar{y}}(t,y) + \dfrac{1}{2}\sigma_{Y}^2 \partial_{yy} h^{\bar{y}}(t,y) \\
& +\left( r+ \widehat{\pi}\left(\mu_S-r + \rho \dfrac{\sigma_S}{\sigma_{Y}} \left( \dfrac{\bar{y}-y}{T-t}-\mu_Y \right) \right)   - \dfrac{1}{2} \widehat{\pi}^2 \sigma_S^2  \gamma(\bar{y}) \right) \big(1-\gamma(\bar{y})\big)  h^{\bar{y}}(t,y).
    \end{split}
\end{equation*}
Finally, using $\gamma(\bar{y}) = \exp(\bar{y})$, we obtain the equilibrium policy
\begin{equation*} 
\begin{split}
     \widehat{\pi}(t,y) & = \dfrac{1}{\sigma_S^2 \mathbb{E}_{t,y}\left[\gamma(Y_T)\right]} \left( \mu_S - r + \rho \sigma_S \sigma_{Y} \int_\mathcal{Y} \dfrac{\partial_y h^{\bar{y}}(t,y) }{h^{\bar{y}}(t,y)}f_{Y_T}(\bar{y}; t,y)d\bar{y} \right) \\
    & = \dfrac{\mu_S - r + \rho \sigma_S \sigma_{Y} \int_\mathcal{Y} \dfrac{\partial_y h^{\bar{y}}(t,y) }{h^{\bar{y}}(t,y)}f_{Y_T}(\bar{y}; t,y)d\bar{y}}{\sigma_S^2 \exp\left(y + \mu_Y(T-t) + \dfrac{1}{2}\sigma_{Y}^{2}(T-t)\right)},
  \end{split}  
\end{equation*}  
which is independent of the current wealth level.
\qed

\clearpage
\section{Additional numerical results} \label{app:additiona_numerics}
In this appendix, we provide supplementary numerics that complement the analysis in Section \ref{sec:application}. 

\begin{figure}[!ht]
\hspace{-3em}
\subfloat[$\mu_Y = 0.02$]{%
    \includegraphics[width=0.6\textwidth]{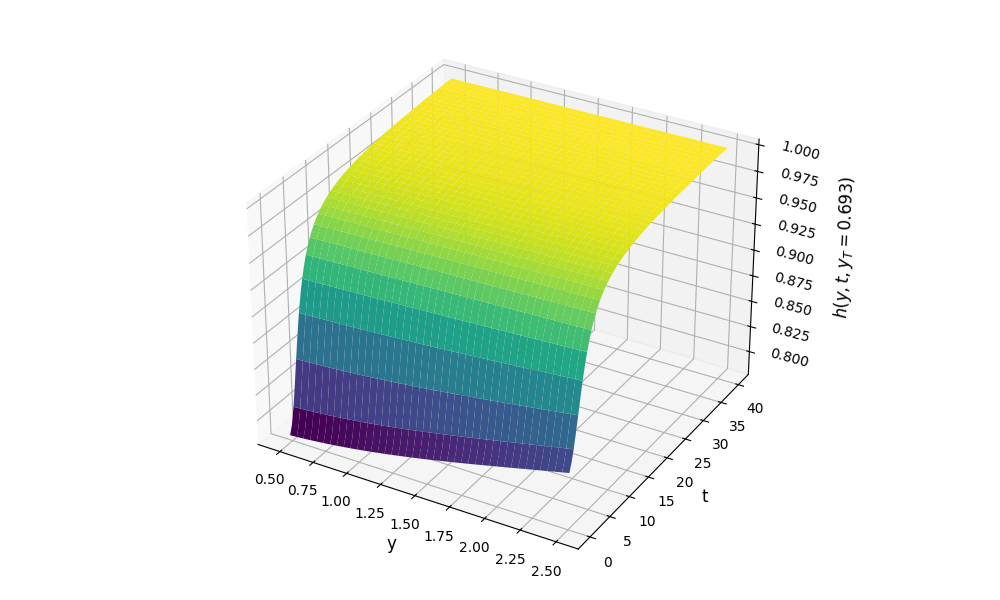}
}
\hspace{-4em}
\subfloat[$\mu_Y = -0.02$]{%
    \includegraphics[width=0.6\textwidth]{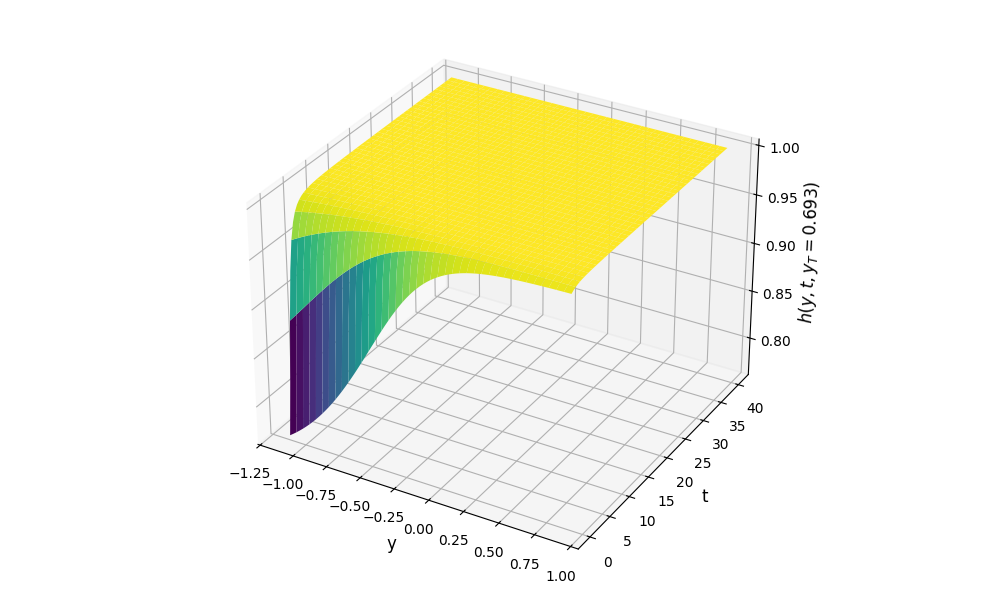}
}
\caption{\footnotesize Intertemporal continuation value $h^{\bar{y}}(t,y)$ for  $\mu_Y = 0.02$ (left) and $\mu_Y = -0.02$ (right), fixing $y_T = \ln(2)$. Other parameters: $(r,\mu_S,\sigma_S,\sigma_Y,\rho,\exp(y_0))=(0.02,0.07,0.2,0.04,0.6,2)$.}
\label{fig:hfunction_continuationvalue}
\end{figure}

First, Figures \ref{fig:hfunction_continuationvalue} (a)-(b) display the functions $h^{\bar{y}}(t,y)$ solving the PIDE \eqref{eq:PIDE_finalform_main} for a fixed value of $\bar{y} = Y_T = \ln(2)$.  

\begin{figure}[!ht]
\centering
{
\includegraphics[width=0.8\textwidth]{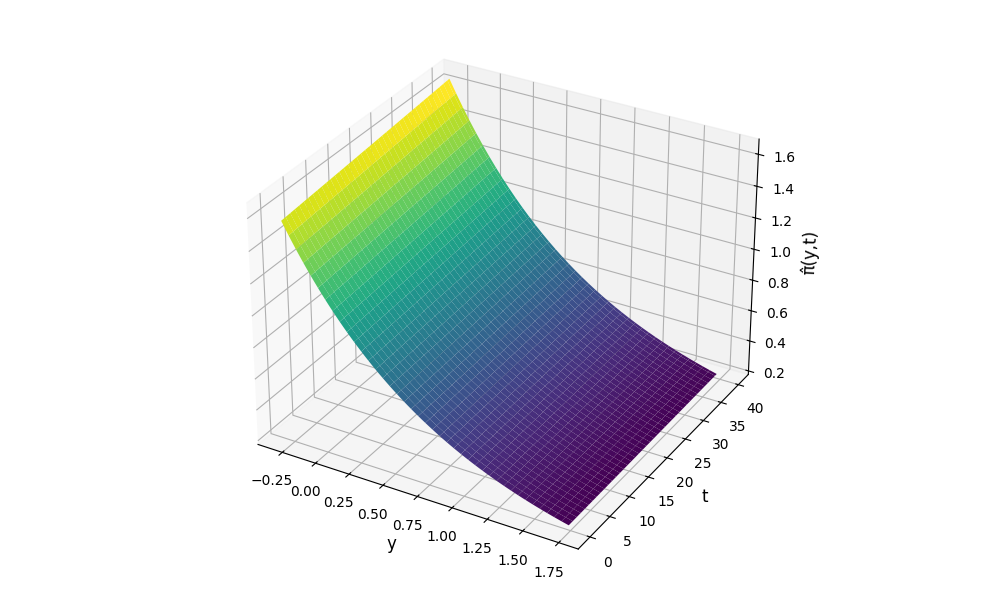}
}
\caption{\footnotesize Equilibrium policy for $\mu_Y = 0.5\sigma_Y^2$. Other parameters: $(r,\mu_S,\sigma_S,\sigma_Y,\rho)=(0.02,0.07,0.2,0.03,0.6)$.}
\label{fig:eqpolicy_cornercase}
\end{figure}

\medskip
Second, Figure \ref{fig:eqpolicy_cornercase} shows the equilibrium investment policy in the special case $\mu_Y = 0.5\sigma_Y^2$. Under this specification, the exponential factor in the denominator of \eqref{eq:equilibrium_policy_semiexplicit_main} becomes time-invariant, leading to an equilibrium policy that is static in $t$.

To conclude, Table \ref{tab:policy_values} reports equilibrium allocations for a range of parameter combinations in $\mu_Y$ and $\rho$. All remaining parameters are as specified above.
\clearpage 
\begin{longtable}{|c|c|cccccc|}
\caption{Equilibrium policy $\hat\pi(y,t)$ for selected values of $(y,t)$. 
Other parameters: $(r,\mu_S,\sigma_S,\sigma_Y,\gamma(y_0)) = (0.02,0.07,0.2,0.04, 2)$.}
\label{tab:policy_values}\\

\toprule
$\mu_Y, \rho$ & $\exp(y)$ & $t=0$ & $t=7$ & $t=14$ & $t=21$ & $t=28$ & $t=35$ \\
\midrule
\endfirsthead

\multicolumn{8}{c}%
{{Table \thetable\ (continued)}} \\
\toprule
$\mu_Y, \rho$ & $\exp(y)$ & $t=0$ & $t=7$ & $t=14$ & $t=21$ & $t=28$ & $t=35$ \\
\midrule
\endhead

\midrule
\multicolumn{8}{r}{{continued on next page}} \\
\endfoot

\bottomrule
\endlastfoot

\multirow{5}{*}{$0.02, 0.6$} 
& 2  & 0.272 & 0.315 & 0.364 & 0.421 & 0.487 & 0.563 \\
& 3  & 0.182 & 0.21 & 0.243 & 0.281 & 0.325 & 0.376 \\
& 4  & 0.136 & 0.157 & 0.182 & 0.211 & 0.244 & 0.282 \\
& 7  & 0.078 & 0.09 & 0.104 & 0.120 & 0.139 & 0.161 \\
& 10  & 0.054 & 0.063 & 0.073 & 0.084 & 0.097 & 0.113 \\
\midrule

\multirow{5}{*}{$0.02, -0.6$} 
& 2  & 0.272 & 0.315 & 0.364 & 0.421 & 0.487 & 0.563 \\
& 3  & 0.181 & 0.21 & 0.243 & 0.281 & 0.325 & 0.376 \\
& 4  & 0.136 & 0.157 & 0.182 & 0.21 & 0.243 & 0.281 \\
& 7  & 0.078 & 0.09 & 0.104 & 0.12 & 0.139 & 0.161 \\
& 10  & 0.054 & 0.063 & 0.073 & 0.084 & 0.097 & 0.113 \\
\midrule

\multirow{5}{*}{$-0.02, 0.6$} 
& 0.8  & 3.48 & 2.946 & 2.574 & 2.251 & 1.967 & 1.72 \\
& 1.2  & 2.328 & 1.963 & 1.716 & 1.5 & 1.312 & 1.147 \\
& 1.6  & 1.724 & 1.472 & 1.287 & 1.125 & 0.984 & 0.86 \\
& 2  & 1.365 & 1.178 & 1.03 & 0.9 & 0.787 & 0.688 \\
& 2.4  & 1.13 & 0.981 & 0.858 & 0.75 & 0.656 & 0.573 \\
\midrule

\multirow{5}{*}{$-0.02, -0.6$} 
& 0.8  & 3.384 & 2.93 & 2.573 & 2.25 & 1.967 & 1.72 \\
& 1.2  & 2.257 & 1.948 & 1.716 & 1.5 & 1.312 & 1.147 \\
& 1.6  & 1.692 & 1.462 & 1.287 & 1.125 & 0.984 & 0.86 \\
& 2  & 1.353 & 1.171 & 1.029 & 0.9 & 0.787 & 0.688 \\
& 2.4  & 1.128 & 0.976 & 0.858 & 0.75 & 0.656 & 0.573 \\
\midrule

\multirow{5}{*}{$0.5\sigma_Y^2, 0.6$} 
& 1 &  1.172 & 1.186 & 1.199 &  1.213 & 1.226 & 1.24   \\
& 2 &   0.586  & 0.593 & 0.599 & 0.606 & 0.613 & 0.62\\
& 3 & 0.391 & 0.395 &  0.4 & 0.404 & 0.409 & 0.413 \\
& 4 &   0.293  & 0.296 & 0.3 & 0.303 & 0.307 & 0.31\\
& 5 &  0.235   & 0.237 & 0.24 & 0.243 & 0.245 & 0.248 \\
\midrule

\multirow{5}{*}{$0.5\sigma_Y^2, -0.6$} 
& 1 & 1.171  & 1.186 & 1.199  & 1.213 & 1.227  &  1.24 \\
& 2 &   0.586  & 0.593 & 0.599 & 0.606 & 0.613 & 0.62\\
& 3 & 0.391 & 0.395 &  0.4 & 0.404 & 0.409 & 0.413 \\
& 4 &   0.293  & 0.296 & 0.3 & 0.303 & 0.307 & 0.31\\
& 5 &  0.234   & 0.237 & 0.24 & 0.243 & 0.245 & 0.248 \\
\midrule

\pagebreak

\multirow{5}{*}{$0.02, 1$} 
& 2  & 0.272 & 0.315 & 0.364 & 0.421 & 0.487 & 0.563\\
&  3 & 0.181 & 0.21 & 0.243 & 0.281 & 0.325 & 0.376 \\
&  4 & 0.136 & 0.157 & 0.182 & 0.21 & 0.243 & 0.282 \\
&  7 & 0.078 & 0.09 & 0.104 & 0.12 & 0.139 & 0.161 \\
&  10 & 0.054 & 0.063 & 0.073  & 0.084 & 0.097 & 0.113\\
\midrule

\multirow{5}{*}{$0.02, -1$} 
& 2  & 0.272 & 0.315 & 0.364 & 0.421 & 0.487 & 0.563\\
&  3 & 0.181 & 0.21 & 0.243 & 0.281 & 0.325 & 0.376 \\
&  4 & 0.136 & 0.157 & 0.182 & 0.21 & 0.243 & 0.282 \\
&  7 & 0.078 & 0.09 & 0.104 & 0.12 & 0.139 & 0.161 \\
&  10 & 0.054 & 0.063 & 0.073  & 0.084 & 0.097 & 0.113\\
\midrule

\multirow{5}{*}{$-0.02, 1$} 
& 0.8  & 3.446 & 3.027 &  2.578 & 2.251 & 1.967 & 1.72\\
& 1.2  & 2.34 & 2.008 &  1.717 & 1.5 & 1.312 & 1.467 \\
&  1.6 & 1.767 & 1.43 &  1.287 & 1.125 & 0.984 & 0.86\\
&  2 & 1.419  & 1.188 & 1.03  & 0.9 & 0.787 & 0.688 \\
& 2.4  & 1.185 & 0.986 &  0.858 & 0.75 & 0.656 & 0.573\\
\midrule

\multirow{5}{*}{$-0.02, -1$} 
&  0.8 & 3.49 & 2.942 & 2.574  & 2.251 & 1.968 & 1.72\\
&  1.2 & 2.387 & 1.962 & 1.716  & 1.5 & 1.312 & 1.147\\
&  1.6  & 1.794 & 1.471 & 1.287  & 1.125 & 0.984 & 0.86\\
&  2 & 1.424 & 1.177 & 1.03 & 0.9 & 0.787 & 0.688 \\
&  2.4 & 1.174 & 0.981 & 0.858 & 0.75 & 0.656 & 0.573\\
\midrule

\multirow{5}{*}{$0.02, 0$} 
& 2  & 0.272 & 0.315 & 0.3634 & 0.421 & 0.487 & 0.563\\
&  3 & 0.181 & 0.21 & 0.243 & 0.281 & 0.325 & 0.376 \\
&  4 & 0.136 & 0.157 & 0.182 & 0.21 & 0.243 & 0.282 \\
&  7 & 0.078 & 0.09 & 0.104 & 0.12 & 0.139 & 0.161 \\
&  10 & 0.054 & 0.063 & 0.073  & 0.084 & 0.097 & 0.113\\
\midrule
\multirow{5}{*}{$-0.02, 0$} 
&  0.8 & 3.368 & 2.946 &  2.574 & 2.25 & 1.968 & 1.72 \\
& 1.2  & 2.245 & 1.963 & 1.716  & 1.5 & 1.312 & 1.147 \\
& 1.6 & 1.684 & 1.472 & 1.287  & 1.125 & 0.984 & 0.86 \\
&  2 & 1.347 & 1.178 & 1.03 & 0.9 & 0.787 & 0.688 \\
& 2.4  & 1.123 & 0.981 &  0.858 & 0.75 & 0.656 & 0.573 \\
\midrule
\end{longtable}

\clearpage
\section{Pseudocodes} \label{app:pseudocode}

To compute the equilibrium policy numerically, we solve the coupled system \eqref{eq:equilibrium_policy_semiexplicit_main}-\eqref{eq:PIDE_finalform_main} using a neural network-based approach. The key idea is to approximate each function $h^{\bar{y}}(t,y)$ by a neural network and to enforce the PIDE through a physics-informed loss that penalizes deviations from the differential equation, the boundary condition, and the coupling with $\epi(t,y)$. The equilibrium policy is updated iteratively: given an estimate of $h^{\bar{y}}$, we compute $\epi$; this updated policy is then fed back into the PIDE, and the networks are trained until a fixed point is reached. 

\bigskip

\SetKwFunction{TrainHModel}{TrainHModel}
\SetKwFunction{PiHat}{EvaluatePiHat}
\SetKwInOut{Input}{Input}
\SetKwInOut{Output}{Output}

\begin{algorithm}[H] \label{algo:NN_pi_hat}
\caption{Evaluate equilibrium policy \( \widehat{\pi}(t, y) \)}\label{alg:pi_hat}
\PiHat{$h_\theta$, current state $y$, time $t$} \\
\Input{Trained model \( h_\theta \), grid size $N_{grid}$ for $y_T$ integration}
\Output{Estimated policy \( \widehat{\pi}(t, y)  \)}

Construct grid \( \{ y_T^{(j)} \}_{j=1}^{N_{grid}} \) over support of $y_T$\;
\For{$j = 1$ \KwTo $N_{grid}$}{Compute \( h_j = h_\theta(t, y, y_T^{(j)}) \)\;
Compute partial derivative $\partial_y h_j(t, y, y_T^{(j)}) \; ; $

Compute ratio:
\[
r_j = \frac{ \partial_y h_j }{ h_j + \varepsilon } \; ;
\]

Using the conditional distribution of the arithmetic Brownian motion, \( Y_T \, \vert \, Y_t = y \sim \mathcal{N}\big(y + \mu_Y (T-t), \sigma_{Y}^2 (T-t)\big) \), compute CDF weights:
\[
w_j = \Phi\left( \frac{y_T^{(j)} - y - \mu_Y (T-t)}{\sigma_{Y} \sqrt{T-t}} \right) \; .
\]
}

Compute integral using trapezoidal rule:
\[
I = \sum_{j=1}^{N_{grid}} r_j \cdot w_j \cdot \Delta y_T \; ;
\]

Return:
\[
\widehat{\pi}(t, y)  = \frac{ \mu_S - r + \rho \sigma_S \sigma_{Y} I }{ \sigma_S^2 \, \mathbb{E}_{t,y}\left[\gamma(y_T)\right] } =  \frac{ \mu_S - r + \rho \sigma_S \sigma_{Y} I }{ \sigma_S^2 \, \exp\left(y + \mu_Y(T-t) + \frac{1}{2}\sigma_{Y}^2 (T-t) \right)}.
\]
\end{algorithm}

\begin{algorithm}[H] \label{algo:NN_PDEsolution}
\caption{Training the NN solution \( h(t, y, y_T) \)}\label{alg:train_h}
\TrainHModel{$\theta$ (NN parameters), terminal time $T$, initial state $Y_0$} \\
\Input{Learning rate $\eta$, number of iterations $N_{\text{iter}}$, batch size $N_{batch}$, sample size $N_{paths}$, boundary loss weight $\lambda_{bc}$}
\Output{Trained model \( h_\theta(t, y, y_T) \)}

Initialize neural network \( h_\theta \colon (t, y, y_T) \mapsto \mathbb{R}_+ \)\;
\For{$k = 1$ \KwTo $N_{iter}$}{
Sample training batch \( \{(t_i, y_i, y_T^{(i)})\}_{i=1}^{N_{batch}} \) from training domain\;
\For{$i = 1$ \KwTo $N_{batch}$}{  
Compute model output \( h_i = h_\theta(t_i, y_i, y_T^{(i)}) \)\;
Compute partial derivatives $
\partial_t h_i, \, \partial_y h_i, \, \partial_{yy} h_i$ with autograd\;

Compute $\pi = \widehat{\pi}(y_i, t_i)$ using  Algorithm \ref{algo:NN_pi_hat}\;

Compute PIDE coefficient functions $P_i = P(t_i, y_i,  y_T^{(i)}), Q_i = Q(t_i, y_i,  y_T^{(i)}),$ $ R_i = R(t_i, y_i, y_T^{(i)}) $:
\begin{eqnarray*}
P_i &=& \left( r + \pi \left( \mu_S - r + \rho \dfrac{\sigma_S}{\sigma_{Y}} \left( \dfrac{y_T^{(i)} - y_i}{T - t_i} - \mu_Y \right) \right) - \dfrac{1}{2} \pi^2 \sigma_S^2 \gamma(y_T^{(i)}) \right)(1 - \gamma(y_T^{(i)})) \;,  \\
Q_i &=& \dfrac{y_T^{(i)} - y_i}{T - t_i} +\rho \sigma_S \sigma_{Y} (1 - \gamma(y_T^{(i)})) \;, \quad R_i =\dfrac{1}{2} \sigma_{Y}^2  \; ;
\end{eqnarray*}

Evaluate PIDE residual:
\[
\text{residual}_i = \partial_t h_i + P(t_i, y_i,  y_T^{(i)}) h_i + Q(t_i, y_i,y_T^{(i)}) \partial_y h_i + R(t_i, y_i, y_T^{(i)}) \partial_{yy} h_i  \; ;
\] 
}

Compute PIDE loss:
\[
L_{pide} = \frac{1}{N_{batch}} \sum_{i=1}^{N_{batch}}(\text{residual}_i)^2 \; ;
\]

Sample boundary points \( \{(T, y_T^{(j)}, y_T^{(j)})\}_{j=1}^{N_{paths}} \)\;
Compute terminal condition loss:
\[
L_{\text{bc}} = \frac{1}{N_{paths}} \sum_{j=1}^{N_{paths}} \left( h_\theta(T, y_T^{(j)}, y_T^{(j)}) - 1 \right)^2 \; ;
\]

Compute total loss: $\quad L = L_{pide} + \lambda_{\text{bc}} L_{\text{bc}} ;$

Update parameters: $\quad \theta \leftarrow \theta - \eta \nabla_\theta L \; ; $
}
\Return \( h_\theta \;. \)
\end{algorithm}

\end{document}